# Hybrid data regression modelling in measurement

Vladimir B. Bokov[*]

NPP Automatica JSC, Russia

**Summery**. Measurement involves the determination of quantitative estimates of physical quantities from experiment, along with estimates of their associated uncertainties. Herewith an experimental system model is the key to extracting information from the experimental data. The measurement information obtained depends directly on the quality of the model. With this concern novel regression modelling techniques have been fashioned by data integration from computer-simulation and physical designed experiments. These techniques have allowed attaining the advanced level of model completeness, parsimony, and precision via approximation of the exact unknown model by mathematical product of available theoretical and appropriate empirical functions. The purpose of this approximation is to represent adequately the true model on the considered region of factor space with all advantages of theoretical modelling. This allows a further focus on the measurement science of issue. Pneumatic gauge hybrid data candidate models' building, solving and validation reviled that such adequate models permit to attain minimum discrepancy from empirical evidence.

## 1. Introduction

Computer models or codes are often used to perform computer-simulation experiments before physical experiments are undertaken. The codes may have high-dimensional inputs, which can be scalars or functions, and output may also be multivariate. In this paper interest is focused on a relatively small set of inputs or explanatory variables and on a single response. The design and analysis of computer experiments has evolved as the power of computer has grown. Sacks *et al*. (1989) provided a review of issues for design and technique used in the analysis of response from computer codes. Making a number of computer code runs at various input values is what they call a computer experiment. And the computer models considered here are deterministic; replicate observations from running the code with the same input values will be identical. This lack of random error makes computer experiments different from physical experiments.

Traditional statistical approaches consider computer and physical experiments separately with corresponding separate designs, analyses, and results. A compelling argument can be made that better, more powerful statistical results can be obtained if we simultaneously analyze the combined data of physical and computer designed experiments. The analysis of such combined data permits the unknown coefficients in an assumed overall regression or response surface model to be estimated more precisely, thereby producing a better-fitting response surface model which is crucial in measurement.

Measurement involves the determination of quantitative estimates of physical quantities from experiment, along with estimates of their associated uncertainties. In this endeavour, a mathematical model of measurement system is required in order to extract information from the experimental data (Cox, Forbes, and Harris, 2002). This implies model building; developing a mathematical model of the experimental system in terms of equations involving parameters that describe all the relevant aspects of the system, and model solving; determining estimates of model parameters from the measured data by solving the equations constructed as part of the model. The response of many measurement systems depends on more than one variable and it is important to model the response of such systems as a function of all relevant explanatory variables or factors. Common approaches to modelling multivariate data have been reviewed by Boudjemaa *et al.* (2003) including approaches specific to data on a regular grid (e.g., tensor product polynomial and splines) and more general approaches (e.g., radial basis functions and support vector machines).

[*] E-mail: bvb@uk2.net

Furthermore, it is useful to classify the types of data arising in metrology into three categories: discrete, continuous and hybrid (Cox, Forbes, and Harris, 2002). Discrete data represent the measurement of a single response variable at a finite number of settings of factors. Continuous data define one or more attributes of the system over a continuum, while hybrid data have both discrete and continuous components. Herewith, Reese *et al.* (2004) noted that it is statistically efficient and desirable to fit a single common response surface model that combines the physical experimental data and the computer model output data to express the relationship between the factors and response variable. And in this paper we make use of designed experiment approach for multivariate hybrid data modelling employing data from two sources; computer-simulation and physical designed experiments. In the both types of experiments the same response variable is considered. Moreover, the computed and observed values of the response are considered for the same set of factors and at the same factor levels in the both experiments. Such approach is of particular importance for measurement situations where the system under study is imperfectly understood. With this advance it is possible to achieve the higher level of model precision and uncertainty reduction.

Investigators of measurement systems are often concerned with elucidation of some functional relationship $E(y) = \eta = F(\xi_1, \xi_2, \ldots, \xi_k) = F(\boldsymbol{\xi})$ connecting the expected value of response $E(y) = \eta$ with $k$ quantitative explanatory variables $\xi_1, \xi_2, \ldots, \xi_k$. Usually, the nature of this relationship is approximated by local polynomial and in function based statistical modelling it is convenient not to have to deal with the actual numerical measures of the variables $\xi_1, \xi_2, \ldots, \xi_k$, but to work with coded variables $x_1, x_2, \ldots, x_k$. These variables are convenient linear transformations of the original $\xi_1, \xi_2, \ldots, \xi_k$ and, so expressions containing the $x_1, x_2, \ldots, x_k$ can always be rewritten in terms of the $\xi_1, \xi_2, \ldots, \xi_k$.

In general, a polynomial $\phi(\boldsymbol{x}, \boldsymbol{\theta})$ in coded variables is a linear combination of powers and products of the $x$'s and $\boldsymbol{\theta}$ is a vector of parameters. The polynomial expression of order $s$ can be thought of as a Taylor's series expansion of the true underlying function $F(\boldsymbol{\xi})$ truncated after terms of $s^{th}$ order. The higher the order of the approximating polynomial, the more closely Taylor series can approximate the true function. The smaller the region over which the approximation needs to be done, the better approximation is possible by given order polynomial. And, in practice over limited region of factor space, a polynomial of only first or second order might adequately represent the true function.

On the other hand in measurement it is possible to use continuous modelling (Reader-Harris, *et al.*, 2000; Esward and Wright, 2007) employing theoretical function to represent the response rather than approximate it by polynomial. In this case a useful theoretical representation takes account of the principal features of the relationship under study. The theoretical function arises from a particular measurement process theory. This theory leads to a set of differential equations which solution $f(\boldsymbol{\xi}, \boldsymbol{\varphi})$ is the function in search. This function depends on the vector $\boldsymbol{\xi} = (\xi_1, \xi_2, \ldots, \xi_k)^T$ of explanatory variables and on the vector $\boldsymbol{\varphi}$ of parameters or physical constants and in contrast to the polynomial serves as an approximation for the true relationship $F(\boldsymbol{\xi})$ over the whole region of factor space.

Moreover, continuous theoretical model contributes to our scientific understanding of the measurement process under study, it provides a better basis for extrapolation, and it is parsimonious in the use of parameters and to give better estimates of the response. So, if we knew theoretical function but did not know the values of the parameters in $\boldsymbol{\varphi}$, then it would be best to fit this function directly to data. However, often these parameters are known from previous experience but the continuous theoretical model is still inadequate.

Essentially, all models are approximations but continuous theoretical and discrete empirical models represent extremes (Box and Draper, 1987). The former would be appropriate in the extreme case where a theory was accurately known about the relationship under study, and the latter would be appropriate in the other extreme case, where nothing could be assumed except the functional relationship between $y$ and $\boldsymbol{\xi}$ was locally smooth. However, generally a modelling situation existing for the most real measurement systems is somewhere in between and, as an investigation proceeds and information is gained, the situation can change. So, it is natural to ask if pure statistical modelling can be merged with theoretic computer model to produce improved predictions; the answer is yes (Bayarri *et al.*, 2007).

Present paper discusses how to reduce modelling uncertainty for real measurement system. When system model is formulated and fitted relying on theoretical considerations only, inferences made from it will be biased and overoptimistic because they ignore missing knowledge and data processing actions which should precede the inference. Thus, the main message of this paper is that in measurement system analysis we can cope with model uncertainty by employing hybrid data regression modelling. For that Section 2 introduces hybrid data multiple regression modelling. Section 3 describes the distributional properties of hybrid data multiple regression. Section 4 provides an example of data analysis comparing multiple linear regression modelling to hybrid data modelling and demonstrates how hybrid data multiple regression techniques may be entertained for considered pneumatic gauge example. Finally, Section 5 provides conclusions.

## 2. Hybrid data multiple regression modelling

Hybrid data regression modelling aims at obtaining overall response surface model, which for measurement system is related to the production of more relevant information of increased utility. Hybrid data model can represent a measurement process in greater detail and with less uncertainty than what is obtainable separately from continuous theoretical or discrete empirical models. Hybrid data modelling is a procedure whereby knowledge and data are fused into consistent, accurate, and intelligible whole. This is concerned with data modelling from two sources where the data come from computer-simulation and physical designed experiments. Many model building and solving problems in measurement involve hybrid data modelling. Until recently, these problems have been addressed using standard methods that do not make the best use of all the data, possibly leading to biases and statistical inefficiency in results. Hybrid data modelling techniques that improve upon existing methods need to be developed, validated, and implemented in the context of and related to integrated model building and solving procedure.

Hybrid data modelling refers to the application of techniques to combine data from computer- simulation and empirical sources with which the quality of information about a known type of behaviour is to be improved. Employing this approach the modelling of measurement process can be enhanced if the exact unknown functional relationship between response and a set of explanatory variables approximate simultaneously by theoretical function $f(\boldsymbol{\xi}, \boldsymbol{\varphi})$, which encapsulates process theoretical knowledge, and proper empirical function $\phi(\boldsymbol{x}, \boldsymbol{\theta})$, which locally approximate a discrepancy between theoretical and empirical data over the considered region of factor space. At this point, model functional structure is formulated by means of these two functions and the model may be written

$$y = \Phi\{f(\boldsymbol{\xi}, \boldsymbol{\varphi}), \phi(\boldsymbol{x}, \boldsymbol{\theta})\} + \delta(\boldsymbol{\xi}) + \varepsilon, \quad (1)$$

where systematic error $\delta(\boldsymbol{\xi}) = \eta - \Phi\{f(\boldsymbol{\xi}, \boldsymbol{\varphi}), \phi(\boldsymbol{x}, \boldsymbol{\theta})\}$ is a difference between the expected value of the response, $E(y) = \eta$, and the value of model functional structure $\Phi\{f(\boldsymbol{\xi}, \boldsymbol{\varphi}), \phi(\boldsymbol{x}, \boldsymbol{\theta})\}$, and $\varepsilon$ is a random error. The formulation of model functional structure may be accomplished in different ways (Bokov, 2007) and expressions discussed in this paper are limited to those based on function multiplication. Therefore, here theoretic-empirical model is written

$$y = f(\boldsymbol{\xi}, \boldsymbol{\varphi}) \times \phi(\boldsymbol{x}, \boldsymbol{\theta}) + e, \quad (2)$$

where $e$ is an experimental error.

This model involves $p$+1 parameters $\boldsymbol{\theta} = (\theta_0, \theta_1 \ldots, \theta_p)^{\mathrm{T}}$ and two important questions that arise are: does the postulated model completely and adequately represent the empirical data and if so, what are the best estimates of the parameters in $\boldsymbol{\theta}$? With this concern we introduce hybrid data multiple regression and for fitting model to data we employ the method of least squares.

### *2.1. Hybrid data model building*

Suppose model (2) functional structure has been approximated by $i$ equations ($i$=1, 2,…, $n$)

$$E(y_i) = f(\xi_{i1}, \xi_{i2}, \ldots, \xi_{ik}; \boldsymbol{\varphi}) \times \{\theta_0 + \theta_1\phi_{i1}(\boldsymbol{x}) + \theta_2\phi_{i2}(\boldsymbol{x}) + \ldots + \theta_p\,\phi_{ip}(\boldsymbol{x})\}, \quad (3)$$

where $f(\xi_{i1}, \xi_{i2}, \ldots, \xi_{ik}; \boldsymbol{\varphi})$ is the representation of theoretical function computation for the $i^{\mathrm{th}}$ combination of actual factor levels $\xi_{i1}, \xi_{i2}, \ldots, \xi_{ik}$ and $\theta_0 + \theta_1\phi_{i1}(\boldsymbol{x}) + \theta_2\phi_{i2}(\boldsymbol{x}) + \ldots + \theta_p\,\phi_{ip}(\boldsymbol{x})$ is the empirical function polynomial approximation for the $i^{\mathrm{th}}$ linear combination of basis functions $\phi_{i1}(\boldsymbol{x}), \phi_{i2}(\boldsymbol{x}), \ldots, \phi_{ip}(\boldsymbol{x})$ which are powers and products of the coded variables $\boldsymbol{x} = (x_1, x_2, \ldots, x_k)^{\mathrm{T}}$. For this approximate model the experimental conditions $(\xi_{i1}, \xi_{i2}, \ldots, \xi_{ik})$ have been run randomly, yielding observations $y_1, y_2, \ldots, y_n$ and for the same conditions the corresponding deterministic computer simulations have been carried out, yielding computations

$$\begin{aligned} z_1 &= f(\xi_{11}, \xi_{12}, \ldots, \xi_{1k}, \boldsymbol{\varphi}),\\ z_2 &= f(\xi_{21}, \xi_{22}, \ldots, \xi_{2k}, \boldsymbol{\varphi}),\\ &\ldots, \\ z_n &= f(\xi_{n1}, \xi_{n2}, \ldots, \xi_{nk}, \boldsymbol{\varphi}). \end{aligned} \quad (4)$$

By means of these simulations we employ the computer-simulation experiment (Sacks *et al.*, 1989) using the theoretical function and randomization theory of experimental design. Therefore, employing the results of computer-simulation and physical designed experiments we obtain a model which relates the observations $y_i$ to the known $z_i$, $\phi_{i1}(\boldsymbol{x}), \phi_{i2}(\boldsymbol{x}), \ldots, \phi_{ip}(\boldsymbol{x})$ and unknown $\theta_0, \theta_1 \ldots, \theta_p$ by $n$ equations

$$\begin{aligned} y_1 &= z_1 \times [\theta_0 + \theta_1\phi_{11}(\boldsymbol{x}) + \theta_2\phi_{12}(\boldsymbol{x}) + \ldots + \theta_p\phi_{1p}(\boldsymbol{x})] + e_1,\\ y_2 &= z_2 \times [\theta_0 + \theta_1\phi_{21}(\boldsymbol{x}) + \theta_2\phi_{22}(\boldsymbol{x}) + \ldots + \theta_p\,\phi_{2p}(\boldsymbol{x})] + e_2,\\ &\ldots,\\ y_n &= z_n \times [\theta_0 + \theta_1\phi_{n1}(\boldsymbol{x}) + \theta_2\phi_{n2}(\boldsymbol{x}) + \ldots + \theta_p\,\phi_{np}(\boldsymbol{x})] + e_n, \end{aligned} \quad (5)$$

where $e_i$ is the errors. This may be written in matrix notation as

$$\boldsymbol{y} = \mathrm{diag}(\boldsymbol{z})\boldsymbol{X}\boldsymbol{\theta} + \boldsymbol{e} \quad (6)$$

Here, in general, $\boldsymbol{y}$ is a $n \times 1$ vector of experimental observations, $\boldsymbol{z}$ is a $n \times 1$ vector of theoretical function computations for the corresponding combinations of actual factor levels from the experimental design, $\boldsymbol{X}=\begin{bmatrix} 1 & \phi_{11}(\boldsymbol{x}) & \phi_{12}(\boldsymbol{x}) & \cdots & \phi_{1p}(\boldsymbol{x}) \\ 1 & \phi_{21}(\boldsymbol{x}) & \phi_{22}(\boldsymbol{x}) & \cdots & \phi_{2p}(\boldsymbol{x}) \\ 1 & \phi_{31}(\boldsymbol{x}) & \phi_{32}(\boldsymbol{x}) & \cdots & \phi_{3p}(\boldsymbol{x}) \\ \vdots & \vdots & \vdots & \ddots & \vdots \\ 1 & \phi_{n1}(\boldsymbol{x}) & \phi_{n2}(\boldsymbol{x}) & \cdots & \phi_{np}(\boldsymbol{x}) \end{bmatrix}$ is an experimental design $n$ x $(p+1)$ full column matrix of the basis functions of coded factor levels with identity element, $\boldsymbol{\theta}$ is a $(p+1)$ x 1 vector of polynomial coefficients, and $\boldsymbol{e}$ is a $n$ x 1 vector of errors for this model. This is a hybrid data multiple regression model and if the diagonal matrix $\boldsymbol{D}$=diag($\boldsymbol{z}$) equals to identity matrix $\boldsymbol{I}$ then we have well-known multiple linear regression model

$$\boldsymbol{y}=\boldsymbol{X\theta}+\boldsymbol{e}_x, \tag{7}$$

where $\boldsymbol{e}_x$ is a $n$ x 1 vector of errors for this model. Thus, the multiple linear regression model may be regarded as the special case of the hybrid data multiple regression model with such a meaning that no computer-simulation data are involved in model building.

## 2.2. *Hybrid data model solving*

In general, the goodness of the least-squares estimates of the parameters in $\boldsymbol{\theta}$ depends on the nature of error distribution. The least-squares estimates would be appropriate if experimental errors were statistically independent, with constant variance, and normally distributed. The Gauss-Markov theorem states that, for the model (7) with elements of $\boldsymbol{e}_x$ pair-wise uncorrelated and having equal variances $\sigma_x^2$, the least-squares estimators of parameters in $\boldsymbol{\theta}$ have the smallest variances of all linear unbiased estimators of these parameters.

The least squares estimates obtained, assuming the model (7) is true, are

$$\boldsymbol{q}=(\boldsymbol{X}^{\mathrm{T}}\boldsymbol{X})^{-1}\boldsymbol{X}^{\mathrm{T}}\boldsymbol{y} \tag{8}$$

and will in general be biased, since

$$E(\boldsymbol{q})=(\boldsymbol{X}^{\mathrm{T}}\boldsymbol{X})^{-1}\boldsymbol{X}^{\mathrm{T}}E(\boldsymbol{y})=(\boldsymbol{X}^{\mathrm{T}}\boldsymbol{X})^{-1}\boldsymbol{X}^{\mathrm{T}}\boldsymbol{DX\theta}, \tag{9}$$

because the true model which should have been fitted is

$$\boldsymbol{y}=\boldsymbol{DX\theta}+\boldsymbol{e} \tag{10}$$

Here the diagonal matrix $\boldsymbol{D}$ may be presented as a sum of identity matrix $\boldsymbol{I}$ and matrix $\boldsymbol{\Delta}=\boldsymbol{D}-\boldsymbol{I}$ then equation (9) becomes

$$E(\boldsymbol{q})=(\boldsymbol{X}^{\mathrm{T}}\boldsymbol{X})^{-1}\boldsymbol{X}^{\mathrm{T}}E(\boldsymbol{y})=(\boldsymbol{X}^{\mathrm{T}}\boldsymbol{X})^{-1}\boldsymbol{X}^{\mathrm{T}}(\boldsymbol{I}+\boldsymbol{\Delta})\boldsymbol{X\theta}=\boldsymbol{\theta}+\boldsymbol{A\theta}, \tag{11}$$

where $\boldsymbol{A}=(\boldsymbol{X}^{\mathrm{T}}\boldsymbol{X})^{-1}\boldsymbol{X}^{\mathrm{T}}\boldsymbol{Y}$ is the alias matrix and $\boldsymbol{Y}=\boldsymbol{\Delta X}=(\boldsymbol{D}-\boldsymbol{I})\boldsymbol{X}$. The alias matrix will be null only if $\boldsymbol{Y}=\boldsymbol{0}$, when $\boldsymbol{D}=\boldsymbol{I}$.

Furthermore, if we suppose that $p+1$ columns of $\boldsymbol{X}$ split into two sets $\boldsymbol{X}_1$, $\boldsymbol{X}_2$ of $p_1$, $p_2$ columns and vector $\boldsymbol{\theta}$ split into two vectors $\boldsymbol{\theta}_1$ and $\boldsymbol{\theta}_2$, respectively, then the model (10) may be rewritten

$$\boldsymbol{y}=\boldsymbol{D}\,\boldsymbol{X}_1\boldsymbol{\theta}_1+\boldsymbol{D}\,\boldsymbol{X}_2\boldsymbol{\theta}_2+\boldsymbol{e} \tag{12}$$

and, taking into account that $\boldsymbol{D}=\boldsymbol{I}+\boldsymbol{\Delta}$, this model may be presented as

$$\boldsymbol{y}=\boldsymbol{X}_1\boldsymbol{\theta}_1+\boldsymbol{X}_2\boldsymbol{\theta}_2+\boldsymbol{Y}_1\boldsymbol{\theta}_1+\boldsymbol{Y}_2\boldsymbol{\theta}_2+\boldsymbol{e}, \tag{13}$$

where $\boldsymbol{Y}_1=(\boldsymbol{D}-\boldsymbol{I})\boldsymbol{X}_1$ and $\boldsymbol{Y}_2=(\boldsymbol{D}-\boldsymbol{I})\boldsymbol{X}_2$. Here, for the multiple linear regression model (7), as a part of the hybrid data model, the simpler model with function $\boldsymbol{X}_1\boldsymbol{\theta}_1$ might be one which might be adequate and function $\boldsymbol{X}_2\boldsymbol{\theta}_2$ might represent further terms which perhaps would have to be added if the terms of $\boldsymbol{X}_1\boldsymbol{\theta}_1$ were inadequate to represent the response. However, for the hybrid data model (13) there is the other opportunity to overcome the inadequacy of the simpler model with function $\boldsymbol{X}_1\boldsymbol{\theta}_1$ by adding the terms of $\boldsymbol{Y}_1\boldsymbol{\theta}_1$ instead of $\boldsymbol{X}_2\boldsymbol{\theta}_2$. Therefore, hybrid data model adequacy will significantly depend on employed theoretical function.

Moreover, if the terms of $\boldsymbol{X}_1\boldsymbol{\theta}_1$ and $\boldsymbol{Y}_1\boldsymbol{\theta}_1$ were inadequate to represent the response, then the terms of $\boldsymbol{X}_2\boldsymbol{\theta}_2$ and $\boldsymbol{Y}_2\boldsymbol{\theta}_2$ could also be added to represent the response adequately.

Using the terms of the multiple linear regression model explicitly the hybrid data model may be formulated as

$$\boldsymbol{y} = \boldsymbol{X}\boldsymbol{\theta} + \boldsymbol{Y}\boldsymbol{\theta} + \boldsymbol{e} \tag{14}$$

Such a model formulation let us an opportunity to analyse the hybrid data model in comparison with the multiple linear regression model. And, because matrices $\boldsymbol{Y}$ and $\boldsymbol{X}$ have the same number of rows, then for model solving the matrix $\boldsymbol{X}$ may be augmented by $\boldsymbol{Y}$ and the model (14) is written

$$\boldsymbol{y} = [\boldsymbol{X}\ \boldsymbol{Y}]\begin{bmatrix}\boldsymbol{\theta}\\ \boldsymbol{\theta}\end{bmatrix} + \boldsymbol{e} \tag{15}$$

Let us denote matrix $[\boldsymbol{X}\ \boldsymbol{Y}]$ by $\boldsymbol{\Psi}$ and vector $\begin{bmatrix}\boldsymbol{\theta}\\ \boldsymbol{\theta}\end{bmatrix}$ by $\boldsymbol{\beta}$. Therefore, the model we deal with is

$$\boldsymbol{y} = \boldsymbol{\Psi}\boldsymbol{\beta} + \boldsymbol{e}, \tag{16}$$

where $\boldsymbol{\beta}$ is a $2(p+1) \times 1$ vector of parameters, $\boldsymbol{\Psi}$ is a $n \times 2(p+1)$ matrix of known values and $\boldsymbol{e}$ is a vector of random error terms. These terms can be defined as $\boldsymbol{e} = \boldsymbol{y} - E(\boldsymbol{y})$ so that $E(\boldsymbol{e}) = \boldsymbol{0}$ and $E(\boldsymbol{y}) = \boldsymbol{\Psi}\boldsymbol{\beta}$. Every element in $\boldsymbol{e}$ has variance $\sigma^2$ and zero covariance with every other element; i.e., $var(\boldsymbol{e}) = E(\boldsymbol{e}\,\boldsymbol{e}^{\mathrm{T}}) = \boldsymbol{I}\sigma^2$. Thus, we have $\boldsymbol{e} \sim (\boldsymbol{0}, \boldsymbol{I}\sigma^2)$ and $\boldsymbol{y} \sim (\boldsymbol{\Psi}\boldsymbol{\beta}, \boldsymbol{I}\sigma^2)$. The more general cases where $\mathrm{var}(\boldsymbol{e}) = \boldsymbol{V}$, whether matrix $\boldsymbol{V}$ be non-singular or singular, are discussed by Searle (1971).

The normal equations corresponding to the model (16) can be derived by least squares and they turn out to be

$$\boldsymbol{\Psi}^{\mathrm{T}}\boldsymbol{\Psi}\boldsymbol{b} = \boldsymbol{\Psi}^{\mathrm{T}}\boldsymbol{y} \tag{17}$$

They involve $\boldsymbol{\Psi}^{\mathrm{T}}\boldsymbol{\Psi}$ which is square and symmetric. Also, because matrix $\boldsymbol{\Psi}$ may be of less than full column rank, then $\boldsymbol{\Psi}^{\mathrm{T}}\boldsymbol{\Psi}$ may be of less than full rank as well. And whenever $\boldsymbol{\Psi}^{\mathrm{T}}\boldsymbol{\Psi}$ is not of full rank the normal equations (17) cannot be solved with one solution, many solutions available. We use $\boldsymbol{b}$ to denote a solution to (17)

$$\boldsymbol{b} = (\boldsymbol{\Psi}^{\mathrm{T}}\boldsymbol{\Psi})^{-}\boldsymbol{\Psi}^{\mathrm{T}}\boldsymbol{y}, \tag{18}$$

where $(\boldsymbol{\Psi}^{\mathrm{T}}\boldsymbol{\Psi})^{-}$ is a generalised inverse of $\boldsymbol{\Psi}^{\mathrm{T}}\boldsymbol{\Psi}$, and it is only solution and not an estimator of $\boldsymbol{\beta}$.

Suppose that vector $\boldsymbol{b}$ is partitioned as $\begin{bmatrix}\boldsymbol{b}_1\\ \boldsymbol{b}_2\end{bmatrix}$ then the normal equations may be presented as

$$\left[\begin{bmatrix}\boldsymbol{X}^T\\ \boldsymbol{Y}^T\end{bmatrix}\begin{bmatrix}\boldsymbol{X} & \boldsymbol{Y}\end{bmatrix}\right]\begin{bmatrix}\boldsymbol{b}_1\\ \boldsymbol{b}_2\end{bmatrix} = \begin{bmatrix}\boldsymbol{X}^T\boldsymbol{y}\\ \boldsymbol{Y}^T\boldsymbol{y}\end{bmatrix} \tag{19}$$

So the corresponding solution is

$$\begin{bmatrix}\boldsymbol{b}_1\\ \boldsymbol{b}_2\end{bmatrix} = \begin{bmatrix}\boldsymbol{X}^T\boldsymbol{X} & \boldsymbol{X}^T\boldsymbol{Y}\\ \boldsymbol{Y}^T\boldsymbol{X} & \boldsymbol{Y}^T\boldsymbol{Y}\end{bmatrix}^{-}\begin{bmatrix}\boldsymbol{X}^T\boldsymbol{y}\\ \boldsymbol{Y}^T\boldsymbol{y}\end{bmatrix} \tag{20}$$

The generalized inverse of symmetric and singular matrix has been derived by Rohde (1965). On defining $\boldsymbol{G}^{-1} = (\boldsymbol{X}^{\mathrm{T}}\boldsymbol{X})^{-1}$ and $\boldsymbol{Q}^{-}$ as generalized inverse of $\boldsymbol{Q}$, where $\boldsymbol{Q} = \boldsymbol{C} - \boldsymbol{B}^{\mathrm{T}}\boldsymbol{G}^{-1}\boldsymbol{B}$, $\boldsymbol{C} = \boldsymbol{Y}^{\mathrm{T}}\boldsymbol{Y}$ and $\boldsymbol{B} = \boldsymbol{X}^{\mathrm{T}}\boldsymbol{Y}$, then the generalized inverse of partitioned matrix is

$$\begin{bmatrix}\boldsymbol{X}^T\boldsymbol{X} & \boldsymbol{X}^T\boldsymbol{Y}\\ \boldsymbol{Y}^T\boldsymbol{X} & \boldsymbol{Y}^T\boldsymbol{Y}\end{bmatrix}^{-} = \begin{bmatrix}\boldsymbol{G}^{-1} + \boldsymbol{G}^{-1}\boldsymbol{B}\boldsymbol{Q}^{-}\boldsymbol{B}^T\boldsymbol{G}^{-1} & -\boldsymbol{G}^{-1}\boldsymbol{B}\boldsymbol{Q}^{-}\\ -\boldsymbol{Q}^{-}\boldsymbol{B}^T\boldsymbol{G}^{-1} & \boldsymbol{Q}^{-}\end{bmatrix}$$

$$=\begin{bmatrix} (\boldsymbol{X}^T\boldsymbol{X})^{-1}+(\boldsymbol{X}^T\boldsymbol{X})^{-1}\boldsymbol{X}^T\boldsymbol{YQ}^-\boldsymbol{Y}^T\boldsymbol{X}(\boldsymbol{X}^T\boldsymbol{X})^{-1} & -(\boldsymbol{X}^T\boldsymbol{X})^{-1}\boldsymbol{X}^T\boldsymbol{YQ}^- \\ -\boldsymbol{Q}^-\boldsymbol{Y}^T\boldsymbol{X}(\boldsymbol{X}^T\boldsymbol{X})^{-1} & \boldsymbol{Q}^- \end{bmatrix}, \tag{21}$$

where $\boldsymbol{Q}^-=[\boldsymbol{Y}^{\mathrm{T}}\{\boldsymbol{I}-\boldsymbol{X}(\boldsymbol{X}^{\mathrm{T}}\boldsymbol{X})^{-1}\boldsymbol{X}^{\mathrm{T}}\}\boldsymbol{Y}]^-$. Therefore, equation (20) may be rewritten as

$$\begin{bmatrix}\boldsymbol{b}_1\\ \boldsymbol{b}_2\end{bmatrix}=\begin{bmatrix} (\boldsymbol{X}^T\boldsymbol{X})^{-1}+(\boldsymbol{X}^T\boldsymbol{X})^{-1}\boldsymbol{X}^T\boldsymbol{YQ}^-\boldsymbol{Y}^T\boldsymbol{X}(\boldsymbol{X}^T\boldsymbol{X})^{-1} & -(\boldsymbol{X}^T\boldsymbol{X})^{-1}\boldsymbol{X}^T\boldsymbol{YQ}^- \\ -\boldsymbol{Q}^-\boldsymbol{Y}^T\boldsymbol{X}(\boldsymbol{X}^T\boldsymbol{X})^{-1} & \boldsymbol{Q}^- \end{bmatrix}\begin{bmatrix}\boldsymbol{X}^T\boldsymbol{y}\\ \boldsymbol{Y}^T\boldsymbol{y}\end{bmatrix} \tag{22}$$

from which

$$\begin{aligned}\boldsymbol{b}_1 &= \{(\boldsymbol{X}^{\mathrm{T}}\boldsymbol{X})^{-1}\boldsymbol{X}^{\mathrm{T}}+(\boldsymbol{X}^{\mathrm{T}}\boldsymbol{X})^{-1}\boldsymbol{X}^{\mathrm{T}}\boldsymbol{Y}\boldsymbol{Q}^-\boldsymbol{Y}^{\mathrm{T}}\boldsymbol{X}(\boldsymbol{X}^{\mathrm{T}}\boldsymbol{X})^{-1}\boldsymbol{X}^{\mathrm{T}}-(\boldsymbol{X}^{\mathrm{T}}\boldsymbol{X})^{-1}\boldsymbol{X}^{\mathrm{T}}\boldsymbol{Y}\boldsymbol{Q}^-\boldsymbol{Y}^{\mathrm{T}}\}\boldsymbol{y}\\ &= (\boldsymbol{X}^{\mathrm{T}}\boldsymbol{X})^{-1}\boldsymbol{X}^{\mathrm{T}}[\boldsymbol{I}-\boldsymbol{Y}\boldsymbol{Q}^-\boldsymbol{Y}^{\mathrm{T}}\{\boldsymbol{I}-\boldsymbol{X}(\boldsymbol{X}^{\mathrm{T}}\boldsymbol{X})^{-1}\boldsymbol{X}^{\mathrm{T}}\}]\boldsymbol{y}\end{aligned} \tag{23}$$

and

$$\begin{aligned}\boldsymbol{b}_2 &= \{\boldsymbol{Q}^-\boldsymbol{Y}^{\mathrm{T}}-\boldsymbol{Q}^-\boldsymbol{Y}^{\mathrm{T}}\boldsymbol{X}(\boldsymbol{X}^{\mathrm{T}}\boldsymbol{X})^{-1}\boldsymbol{X}^{\mathrm{T}}\}\boldsymbol{y}\\ &= \boldsymbol{Q}^-\boldsymbol{Y}^{\mathrm{T}}\{\boldsymbol{I}-\boldsymbol{X}(\boldsymbol{X}^{\mathrm{T}}\boldsymbol{X})^{-1}\boldsymbol{X}^{\mathrm{T}}\}\boldsymbol{y}\end{aligned} \tag{24}$$

Further, defining

$$\boldsymbol{Z}=\{\boldsymbol{I}-\boldsymbol{X}(\boldsymbol{X}^{\mathrm{T}}\boldsymbol{X})^{-1}\boldsymbol{X}^{\mathrm{T}}\}\boldsymbol{Y} \tag{25}$$

as the matrix of deviations of matrix $\boldsymbol{Y}$ from its calculated values with the use of experimental design matrix and because matrix $\boldsymbol{I}-\boldsymbol{X}(\boldsymbol{X}^{\mathrm{T}}\boldsymbol{X})^{-1}\boldsymbol{X}^{\mathrm{T}}$ is idempotent, hence just as in (18) we can write (24) as

$$\boldsymbol{b}_2=(\boldsymbol{Z}^{\mathrm{T}}\boldsymbol{Z})^-\boldsymbol{Z}^{\mathrm{T}}\boldsymbol{y} \tag{26}$$

Then $\boldsymbol{b}_1$ is given by

$$\boldsymbol{b}_1=(\boldsymbol{X}^{\mathrm{T}}\boldsymbol{X})^{-1}\boldsymbol{X}^{\mathrm{T}}\{\boldsymbol{I}-\boldsymbol{Y}(\boldsymbol{Z}^{\mathrm{T}}\boldsymbol{Z})^-\boldsymbol{Z}^{\mathrm{T}}\}\boldsymbol{y} \tag{27}$$

### *2.3. Consequences of solution for model*

For the hybrid data model the solution $\boldsymbol{b}$ is a function of observations $\boldsymbol{y}$ and for the generalized inverse $(\boldsymbol{\Psi}^{\mathrm{T}}\boldsymbol{\Psi})^-$,

$$E(\boldsymbol{b})=(\boldsymbol{\Psi}^{\mathrm{T}}\boldsymbol{\Psi})^-\boldsymbol{\Psi}^{\mathrm{T}}E(\boldsymbol{y})=(\boldsymbol{\Psi}^{\mathrm{T}}\boldsymbol{\Psi})^-\boldsymbol{\Psi}^{\mathrm{T}}\boldsymbol{\Psi}\boldsymbol{\beta}=\boldsymbol{J}\boldsymbol{\beta}, \tag{28}$$

i.e. $\boldsymbol{b}$ has expected value $\boldsymbol{J}\boldsymbol{\beta}$ where $\boldsymbol{J}=(\boldsymbol{\Psi}^{\mathrm{T}}\boldsymbol{\Psi})^-\boldsymbol{\Psi}^{\mathrm{T}}\boldsymbol{\Psi}$. Hence $\boldsymbol{b}$ is an unbiased estimator of $\boldsymbol{J}\boldsymbol{\beta}$.

The uncertainty or variance-covariance matrix of $\boldsymbol{b}$ is

$$\begin{aligned} var(\boldsymbol{b}) &= var((\boldsymbol{\Psi}^{\mathrm{T}}\boldsymbol{\Psi})^-\boldsymbol{\Psi}^{\mathrm{T}}\boldsymbol{y})\\ &= (\boldsymbol{\Psi}^{\mathrm{T}}\boldsymbol{\Psi})^-\boldsymbol{\Psi}^{\mathrm{T}}var(\boldsymbol{y})\,\boldsymbol{\Psi}\{(\boldsymbol{\Psi}^{\mathrm{T}}\boldsymbol{\Psi})^-\}^{\mathrm{T}}\\ &= (\boldsymbol{\Psi}^{\mathrm{T}}\boldsymbol{\Psi})^-\boldsymbol{\Psi}^{\mathrm{T}}\boldsymbol{\Psi}\{(\boldsymbol{\Psi}^{\mathrm{T}}\boldsymbol{\Psi})^-\}^{\mathrm{T}}\sigma^2 \end{aligned} \tag{29}$$

Thus, the matrix $(\boldsymbol{\Psi}^{\mathrm{T}}\boldsymbol{\Psi})^-\boldsymbol{\Psi}^{\mathrm{T}}\boldsymbol{\Psi}\{(\boldsymbol{\Psi}^{\mathrm{T}}\boldsymbol{\Psi})^-\}^{\mathrm{T}}$ determines the variances and covariances of the elements in $\boldsymbol{b}$.

A similar result holds for $\boldsymbol{b}_2$: using the partitioning form of $(\boldsymbol{\Psi}^{\mathrm{T}}\boldsymbol{\Psi})^-$ shown in (21), with $\boldsymbol{Q}^- = (\boldsymbol{Z}^{\mathrm{T}}\boldsymbol{Z})^-$, result (29) becomes (see Appendix A)

$$var\begin{bmatrix}\boldsymbol{b}_1\\ \boldsymbol{b}_2\end{bmatrix}=\begin{bmatrix}\boldsymbol{G}^{-1}\{\boldsymbol{I}+\boldsymbol{X}^{\mathrm{T}}\boldsymbol{YQ}^-\boldsymbol{YXG}^{-1}-\boldsymbol{X}^{\mathrm{T}}(\boldsymbol{I}-\boldsymbol{YQ}^-\boldsymbol{Z}^{\mathrm{T}})\boldsymbol{YG}^{-1}\boldsymbol{X}^{\mathrm{T}}\boldsymbol{YQ}^-\} & -\boldsymbol{Q}^-\boldsymbol{Y}^{\mathrm{T}}\boldsymbol{XG}^{-1}+\boldsymbol{G}^{-1}\boldsymbol{X}^{\mathrm{T}}(\boldsymbol{I}-\boldsymbol{YQ}^-\boldsymbol{Z}^{\mathrm{T}})\boldsymbol{YQ}^- \\ -\boldsymbol{Q}^-\boldsymbol{Z}^{\mathrm{T}}\boldsymbol{YG}^{-1}\boldsymbol{X}^{\mathrm{T}}\boldsymbol{YQ}^- & \boldsymbol{Q}^-\boldsymbol{Z}^{\mathrm{T}}\boldsymbol{YQ}^-\end{bmatrix}\sigma^2 \tag{30}$$

Hence

$$var(\boldsymbol{b}_2)=\boldsymbol{Q}^-\boldsymbol{Z}^{\mathrm{T}}\boldsymbol{Y}\boldsymbol{Q}^-\sigma^2 \tag{31}$$

and

$$var(\boldsymbol{b}_1)=(\boldsymbol{X}^{\mathrm{T}}\boldsymbol{X})^{-1}\{\boldsymbol{I}+\boldsymbol{X}^{\mathrm{T}}\boldsymbol{Y}\boldsymbol{Q}^-\boldsymbol{Y}^{\mathrm{T}}\boldsymbol{X}(\boldsymbol{X}^{\mathrm{T}}\boldsymbol{X})^{-1}-\boldsymbol{X}^{\mathrm{T}}(\boldsymbol{I}-\boldsymbol{Y}\boldsymbol{Q}^-\boldsymbol{Z}^{\mathrm{T}})\boldsymbol{Y}(\boldsymbol{X}^{\mathrm{T}}\boldsymbol{X})^{-1}\boldsymbol{X}^{\mathrm{T}}\boldsymbol{Y}\boldsymbol{Q}^-\}\sigma^2 \tag{32}$$

plus $cov(\boldsymbol{b}_1, \boldsymbol{b}_2)=-\{\boldsymbol{Q}^-\boldsymbol{Y}^{\mathrm{T}}\boldsymbol{X}(\boldsymbol{X}^{\mathrm{T}}\boldsymbol{X})^{-1}+(\boldsymbol{X}^{\mathrm{T}}\boldsymbol{X})^{-1}\boldsymbol{X}^{\mathrm{T}}(\boldsymbol{I}-\boldsymbol{Y}\boldsymbol{Q}^-\boldsymbol{Z}^{\mathrm{T}})\boldsymbol{Y}\boldsymbol{Q}^-\}\sigma^2.$

### *2.4. Estimating E(y) for hybrid data model*

Corresponding to the vector of observations $\boldsymbol{y}$ we have the vector of estimated expected values

$$\hat{E}(\boldsymbol{y})=\hat{\boldsymbol{y}}=\boldsymbol{\Psi}\boldsymbol{b}=\boldsymbol{\Psi}(\boldsymbol{\Psi}^{\mathrm{T}}\boldsymbol{\Psi})^-\boldsymbol{\Psi}^{\mathrm{T}}\boldsymbol{y}. \tag{33}$$

This vector is invariant to the choice of generalized inverse of $\boldsymbol{\Psi}^{\mathrm{T}}\boldsymbol{\Psi}$ which is used for $(\boldsymbol{\Psi}^{\mathrm{T}}\boldsymbol{\Psi})^{-}$, because matrix $\boldsymbol{\Psi}(\boldsymbol{\Psi}^{\mathrm{T}}\boldsymbol{\Psi})^{-}\boldsymbol{\Psi}^{\mathrm{T}}$ is invariant (Searle, 1971). Hence, $\hat{\boldsymbol{y}}$ is the vector of estimated expected values corresponding to the vector $\boldsymbol{y}$ of observations. This means that no matter what solution of the normal equations is used for $\boldsymbol{b}$ the vector $\hat{\boldsymbol{y}}$ will always be the same. It implies that we can get a solution to the normal equations in any way, call it $\boldsymbol{b}$, and no matter which solution it is, $\boldsymbol{\Psi b}$ will be correct value of $\hat{\boldsymbol{y}}$.

The uncertainty matrix of estimator $\hat{\boldsymbol{y}} = \boldsymbol{\Psi b}$ is readily obtained using variance operator

$$var(\hat{\boldsymbol{y}})=var(\boldsymbol{\Psi b})=\boldsymbol{\Psi} var(\boldsymbol{b})\,\boldsymbol{\Psi}^{\mathrm{T}} = \boldsymbol{\Psi}(\boldsymbol{\Psi}^{\mathrm{T}}\boldsymbol{\Psi})^{-}\boldsymbol{\Psi}^{\mathrm{T}}\boldsymbol{\Psi}\{(\boldsymbol{\Psi}^{\mathrm{T}}\boldsymbol{\Psi})^{-}\}^{\mathrm{T}}\boldsymbol{\Psi}^{\mathrm{T}}\sigma^2=\boldsymbol{\Psi}(\boldsymbol{\Psi}^{\mathrm{T}}\boldsymbol{\Psi})^{-}\boldsymbol{\Psi}^{\mathrm{T}}\sigma^2 \quad (34)$$

On substituting $\boldsymbol{\Psi}= [\boldsymbol{X}\ \boldsymbol{Y}]$ and using $\boldsymbol{Q}^{-}$, this converts (see Appendix B) to

$$var(\hat{\boldsymbol{y}})=\{\boldsymbol{X}(\boldsymbol{X}^{\mathrm{T}}\boldsymbol{X})^{-1}\boldsymbol{X}^{\mathrm{T}}+\boldsymbol{Z}(\boldsymbol{Z}^{\mathrm{T}}\boldsymbol{Z})^{-}\boldsymbol{Z}^{\mathrm{T}}\}\,\sigma^2 \quad (35)$$

### *2.5. Residual sum of squares*

For hybrid data model the vector of fitted values corresponding to the observed values can be obtained from equation (33). The residual sum of squares in this case is

$$\begin{aligned} SS_E &=(\boldsymbol{y}-\hat{\boldsymbol{y}})^{\mathrm{T}}(\boldsymbol{y}-\hat{\boldsymbol{y}}) \\ &=\boldsymbol{y}^{\mathrm{T}}\{\boldsymbol{I}-\boldsymbol{\Psi}(\boldsymbol{\Psi}^{\mathrm{T}}\boldsymbol{\Psi})^{-}\boldsymbol{\Psi}^{\mathrm{T}}\}^{\mathrm{T}}\{\boldsymbol{I}-\boldsymbol{\Psi}(\boldsymbol{\Psi}^{\mathrm{T}}\boldsymbol{\Psi})^{-}\boldsymbol{\Psi}^{\mathrm{T}}\}\boldsymbol{y} \\ &=\boldsymbol{y}^{\mathrm{T}}\{\boldsymbol{I}-\boldsymbol{\Psi}(\boldsymbol{\Psi}^{\mathrm{T}}\boldsymbol{\Psi})^{-}\boldsymbol{\Psi}^{\mathrm{T}}\}\boldsymbol{y} \quad (36) \end{aligned}$$

because matrix $\boldsymbol{I}-\boldsymbol{\Psi}(\boldsymbol{\Psi}^{\mathrm{T}}\boldsymbol{\Psi})^{-}\boldsymbol{\Psi}^{\mathrm{T}}$ is symmetric and idempotent. Further, because $\boldsymbol{\Psi}(\boldsymbol{\Psi}^{\mathrm{T}}\boldsymbol{\Psi})^{-}\boldsymbol{\Psi}^{\mathrm{T}}$ is invariant to $(\boldsymbol{\Psi}^{\mathrm{T}}\boldsymbol{\Psi})^{-}$, so is $SS_E$. Hence, $SS_E$ is invariant to whatever solution of the normal equations is used for $\boldsymbol{b}$. This is another result invariant to the many solutions of the normal equations.

In (36) $SS_E$ is written as a quadratic form in $\boldsymbol{y}$ therefore, with $\boldsymbol{y}$ being distributed $(\boldsymbol{\Psi\beta}, \boldsymbol{I}\sigma^2)$ the expected value of $SS_E$ is

$$\begin{aligned} E(SS_E) &=E[\boldsymbol{y}^{\mathrm{T}}\{\boldsymbol{I}-\boldsymbol{\Psi}(\boldsymbol{\Psi}^{\mathrm{T}}\boldsymbol{\Psi})^{-}\boldsymbol{\Psi}^{\mathrm{T}}\}\boldsymbol{y}] \\ &=\mathrm{tr}[\{\boldsymbol{I}-\boldsymbol{\Psi}(\boldsymbol{\Psi}^{\mathrm{T}}\boldsymbol{\Psi})^{-}\boldsymbol{\Psi}^{\mathrm{T}}\}\boldsymbol{I}\sigma^2]+(\boldsymbol{\Psi\beta})^{\mathrm{T}}\{\boldsymbol{I}-\boldsymbol{\Psi}(\boldsymbol{\Psi}^{\mathrm{T}}\boldsymbol{\Psi})^{-}\boldsymbol{\Psi}^{\mathrm{T}}\}\boldsymbol{\Psi\beta} \quad (37) \end{aligned}$$

Through the properties of $\boldsymbol{\Psi}(\boldsymbol{\Psi}^{\mathrm{T}}\boldsymbol{\Psi})^{-}\boldsymbol{\Psi}^{\mathrm{T}}$ and due to the fact that the trace of an idempotent matrix equals its rank $E(SS_E)=\mathrm{rank}\{\boldsymbol{I}-\boldsymbol{\Psi}(\boldsymbol{\Psi}^{\mathrm{T}}\boldsymbol{\Psi})^{-}\boldsymbol{\Psi}^{\mathrm{T}}\}\sigma^2=[n-\mathrm{rank}\{\boldsymbol{\Psi}(\boldsymbol{\Psi}^{\mathrm{T}}\boldsymbol{\Psi})^{-}\boldsymbol{\Psi}^{\mathrm{T}}\}]\sigma^2=\{n-\mathrm{rank}(\boldsymbol{\Psi})\}\sigma^2$. Hence, an unbiased estimator of $\sigma^2$ is

$$\hat{\sigma}^2 = SS_E / \{n-\mathrm{rank}(\boldsymbol{\Psi})\} \quad (38)$$

The use of rank($\boldsymbol{\Psi}$) in the expectation is because matrix $\boldsymbol{\Psi}$ may be not of full column rank. The rank of $\boldsymbol{\Psi}$ depends on the nature of data available and design matrix $\boldsymbol{X}$ employed. When the choice of design matrix $\boldsymbol{X}$ follows to condition $n \le 2(p+1)$ the rank of $\boldsymbol{\Psi}$ is always equal to $n$. In this case from equation (38) $SS_E=0$ and the estimation of $\sigma^2$ becomes impossible without experimental arrangement in which two or more runs are made at an identical set of levels of the explanatory variables. These runs should be made in such a way that they are subject to all the sources of errors that beset runs made at different sets of levels of the explanatory variables. Thus, employing the experimental designs which give rank($\boldsymbol{\Psi}$)=$n$ the hybrid data model has minimum residual sum of squares which equal to its pure error sum of squares.

An expression for $SS_E$ involving $\boldsymbol{X}$ and $\boldsymbol{Z}$ can also be established. For (36) is equivalent to

$$SS_E=\boldsymbol{y}^{\mathrm{T}}\{\boldsymbol{I}-\boldsymbol{X}(\boldsymbol{X}^{\mathrm{T}}\boldsymbol{X})^{-1}\boldsymbol{X}^{\mathrm{T}}-\boldsymbol{Z}(\boldsymbol{Z}^{\mathrm{T}}\boldsymbol{Z})^{-}\boldsymbol{Z}^{\mathrm{T}}\}\boldsymbol{y} \quad (39)$$

because from (34) and (35) $\boldsymbol{\Psi}(\boldsymbol{\Psi}^{\mathrm{T}}\boldsymbol{\Psi})^{-}\boldsymbol{\Psi}^{\mathrm{T}}=\boldsymbol{X}(\boldsymbol{X}^{\mathrm{T}}\boldsymbol{X})^{-1}\boldsymbol{X}^{\mathrm{T}}+\boldsymbol{Z}(\boldsymbol{Z}^{\mathrm{T}}\boldsymbol{Z})^{-}\boldsymbol{Z}^{\mathrm{T}}$. In equation (39) $SS_E$ is written as a quadratic form in $\boldsymbol{y}$. Therefore, with $\boldsymbol{y}$ being distributed $(\boldsymbol{\Psi\beta}, \boldsymbol{I}\sigma^2)$ then the expected value of $SS_E$ is

$$E(SS_E)=\mathrm{tr}\{\boldsymbol{I}-\boldsymbol{X}(\boldsymbol{X}^{\mathrm{T}}\boldsymbol{X})^{-1}\boldsymbol{X}^{\mathrm{T}}-\boldsymbol{Z}(\boldsymbol{Z}^{\mathrm{T}}\boldsymbol{Z})^{-}\boldsymbol{Z}^{\mathrm{T}}\}\boldsymbol{I}\sigma^2+(\boldsymbol{\Psi\beta})^{\mathrm{T}}\{\boldsymbol{I}-\boldsymbol{\Psi}(\boldsymbol{\Psi}^{\mathrm{T}}\boldsymbol{\Psi})^{-}\boldsymbol{\Psi}^{\mathrm{T}}\}\boldsymbol{\Psi\beta} \quad (40)$$

And due to the fact that the trace of an idempotent matrix equals its rank we have $E(SS_E)=\mathrm{rank}\{\boldsymbol{I}-\boldsymbol{X}(\boldsymbol{X}^{\mathrm{T}}\boldsymbol{X})^{-1}\boldsymbol{X}^{\mathrm{T}}-\boldsymbol{Z}(\boldsymbol{Z}^{\mathrm{T}}\boldsymbol{Z})^{-}\boldsymbol{Z}^{\mathrm{T}}\}\sigma^2=[n-\mathrm{rank}\{\boldsymbol{X}(\boldsymbol{X}^{\mathrm{T}}\boldsymbol{X})^{-1}\boldsymbol{X}^{\mathrm{T}}\}-\mathrm{rank}\{\boldsymbol{Z}(\boldsymbol{Z}^{\mathrm{T}}\boldsymbol{Z})^{-}\boldsymbol{Z}^{\mathrm{T}}\}]\sigma^2 = \{n$

–rank($\boldsymbol{X}$)–rank($\boldsymbol{Z}$)}$\sigma^2$. Hence, because rank($\boldsymbol{X}$) $=p+1$ then an unbiased estimate of $\sigma^2$ is $\hat{\sigma}^2 = SS_E / \{n–p–1–\text{rank}(\boldsymbol{Z})\}$ which is the same as (38) because rank($\boldsymbol{\Psi}$)= rank($\boldsymbol{X}$)+rank($\boldsymbol{Z}$).

### 2.6. *Partitioning the total sum of squares*

The total sum of squares is $SS_T = \boldsymbol{y}^{\mathrm{T}}\boldsymbol{y}$ and the residual sum of squares for the hybrid data model has been defined by equation (36). The difference

$$SS_R = SS_T - SS_E = \boldsymbol{y}^{\mathrm{T}}\boldsymbol{\Psi}(\boldsymbol{\Psi}^{\mathrm{T}}\boldsymbol{\Psi})^{-}\boldsymbol{\Psi}^{\mathrm{T}}\boldsymbol{y} = \boldsymbol{b}^{\mathrm{T}}\boldsymbol{\Psi}^{\mathrm{T}}\boldsymbol{y} \tag{41}$$

represents the portion of $SS_T$ attributable to having fitted the regression and called the sum of squares due to regression. And with regard to equation (39) the regression sum of squares for hybrid data model may be rewritten as

$$SS_R = \boldsymbol{y}^{\mathrm{T}}\boldsymbol{X}(\boldsymbol{X}^{\mathrm{T}}\boldsymbol{X})^{-1}\boldsymbol{X}^{\mathrm{T}}\boldsymbol{y} + \boldsymbol{b}_2^{\mathrm{T}}\boldsymbol{Z}^{\mathrm{T}}\boldsymbol{y}, \tag{42}$$

because $\boldsymbol{b}_2^{\mathrm{T}} = \boldsymbol{y}^{\mathrm{T}}\boldsymbol{Z}(\boldsymbol{Z}^{\mathrm{T}}\boldsymbol{Z})^{-}$. Then we see that

$$SS_R = SS_{Rx} + SS_{Rc} \tag{43}$$

where $SS_{Rx} = \boldsymbol{y}^{\mathrm{T}}\boldsymbol{X}(\boldsymbol{X}^{\mathrm{T}}\boldsymbol{X})^{-1}\boldsymbol{X}^{\mathrm{T}}\boldsymbol{y}$ and, so we can call

$$SS_{Rc} = SS_R - SS_{Rx} = \boldsymbol{b}_2^{\mathrm{T}}\boldsymbol{Z}^{\mathrm{T}}\boldsymbol{y} \tag{44}$$

the hybrid data regression sum of squares corrected for the sum of squares due to multiple linear regression. Similar to $SS_{Rc}$ the corrected sum of squares of the $y$'s is

$$SS_{Tc} = SS_T - SS_{Rx} = \boldsymbol{y}^{\mathrm{T}}\boldsymbol{y} - \boldsymbol{y}^{\mathrm{T}}\boldsymbol{X}(\boldsymbol{X}^{\mathrm{T}}\boldsymbol{X})^{-1}\boldsymbol{X}^{\mathrm{T}}\boldsymbol{y} = \boldsymbol{y}^{\mathrm{T}}\{\boldsymbol{I} - \boldsymbol{X}(\boldsymbol{X}^{\mathrm{T}}\boldsymbol{X})^{-1}\boldsymbol{X}^{\mathrm{T}}\}\boldsymbol{y} = \boldsymbol{\upsilon}^{\mathrm{T}}\boldsymbol{\upsilon}, \tag{45}$$

where $\boldsymbol{\upsilon} = \{\boldsymbol{I} - \boldsymbol{X}(\boldsymbol{X}^{\mathrm{T}}\boldsymbol{X})^{-1}\boldsymbol{X}^{\mathrm{T}}\}\boldsymbol{y}$.

The three forms of partitioning the sums of squares are shown in Table 1. The first column shows the sums of squares attributable to fitting the model (16). In the second column $SS_{Rx}$ is the sum of squares due to multiple linear regression and $SS_{Rc}$ is the sum of squares for fitting the model, corrected for the sum of squares due to multiple linear regression. The third column is identical to the second except that $SS_{Rx}$ has been deleted from the body of the table and subtracted from $SS_T$ to give $SS_{Tc}$ as the total sum of squares corrected for the sum of squares due to multiple linear regression. Table 1 form the basis of analysis of variance table for hybrid data regression.

Table 1. Partitioning the sums of squares for hybrid data regression.

| | | |
|---|---|---|
| | $SS_{Rx} = \boldsymbol{y}^{\mathrm{T}}\boldsymbol{X}(\boldsymbol{X}^{\mathrm{T}}\boldsymbol{X})^{-1}\boldsymbol{X}^{\mathrm{T}}\boldsymbol{y}$ | |
| $SS_R = \boldsymbol{y}^{\mathrm{T}}\boldsymbol{\Psi}(\boldsymbol{\Psi}^{\mathrm{T}}\boldsymbol{\Psi})^{-}\boldsymbol{\Psi}^{\mathrm{T}}\boldsymbol{y}$ | $SS_{Rc} = \boldsymbol{y}^{\mathrm{T}}\{\boldsymbol{\Psi}(\boldsymbol{\Psi}^{\mathrm{T}}\boldsymbol{\Psi})^{-}\boldsymbol{\Psi}^{\mathrm{T}} - \boldsymbol{X}(\boldsymbol{X}^{\mathrm{T}}\boldsymbol{X})^{-1}\boldsymbol{X}^{\mathrm{T}}\}\boldsymbol{y}$ | $SS_{Rc} = \boldsymbol{y}^{\mathrm{T}}\{\boldsymbol{\Psi}(\boldsymbol{\Psi}^{\mathrm{T}}\boldsymbol{\Psi})^{-}\boldsymbol{\Psi}^{\mathrm{T}} - \boldsymbol{X}(\boldsymbol{X}^{\mathrm{T}}\boldsymbol{X})^{-1}\boldsymbol{X}^{\mathrm{T}}\}\boldsymbol{y}$ |
| $SS_E = \boldsymbol{y}^{\mathrm{T}}\{\boldsymbol{I} - \boldsymbol{\Psi}(\boldsymbol{\Psi}^{\mathrm{T}}\boldsymbol{\Psi})^{-}\boldsymbol{\Psi}^{\mathrm{T}}\}\boldsymbol{y}$ | $SS_E = \boldsymbol{y}^{\mathrm{T}}\{\boldsymbol{I} - \boldsymbol{\Psi}(\boldsymbol{\Psi}^{\mathrm{T}}\boldsymbol{\Psi})^{-}\boldsymbol{\Psi}^{\mathrm{T}}\}\boldsymbol{y}$ | $SS_E = \boldsymbol{y}^{\mathrm{T}}\{\boldsymbol{I} - \boldsymbol{\Psi}(\boldsymbol{\Psi}^{\mathrm{T}}\boldsymbol{\Psi})^{-}\boldsymbol{\Psi}^{\mathrm{T}}\}\boldsymbol{y}$ |
| $SS_T = \boldsymbol{y}^{\mathrm{T}}\boldsymbol{y}$ | $SS_T = \boldsymbol{y}^{\mathrm{T}}\boldsymbol{y}$ | $SS_{Tc} = \boldsymbol{\upsilon}^{\mathrm{T}}\boldsymbol{\upsilon}$ |

## 3. Distributional properties of multiple hybrid data regression

If errors $\boldsymbol{e}$ are normally distributed $\boldsymbol{e} \sim N(\boldsymbol{0}, \boldsymbol{I}\sigma^2)$ then the distributional properties of $\boldsymbol{y}$ and functions of $\boldsymbol{y}$ follow directly. From (16) we have $\boldsymbol{e} = \boldsymbol{y} - \boldsymbol{\Psi\beta}$ and, therefore $\boldsymbol{y} \sim N(\boldsymbol{\Psi\beta}, \boldsymbol{I}\sigma^2)$. The $\boldsymbol{b}$ solution is a linear function of $\boldsymbol{y}$, hence it is normally distributed

$$\boldsymbol{b} \sim N[\boldsymbol{J\beta}, (\boldsymbol{\Psi}^{\mathrm{T}}\boldsymbol{\Psi})^{-}\boldsymbol{\Psi}^{\mathrm{T}}\boldsymbol{\Psi}\{(\boldsymbol{\Psi}^{\mathrm{T}}\boldsymbol{\Psi})^{-}\}^{\mathrm{T}}\sigma^2].$$

The $\boldsymbol{b}$ is independent from $\hat{\sigma}^2$. To show this we have $\boldsymbol{b}$ as defined by equation (18) and $SS_E$ – by equation (36). So, when $\boldsymbol{y} \sim N(\boldsymbol{\Psi\beta}, \boldsymbol{I}\sigma^2)$, then $SS_E$ and $\boldsymbol{b}$ are distributed independently because $(\boldsymbol{\Psi}^{\mathrm{T}}\boldsymbol{\Psi})^{-}\boldsymbol{\Psi}^{\mathrm{T}}\boldsymbol{I}\sigma^2\{\boldsymbol{I} - \boldsymbol{\Psi}(\boldsymbol{\Psi}^{\mathrm{T}}\boldsymbol{\Psi})^{-}\boldsymbol{\Psi}^{\mathrm{T}}\} = \{(\boldsymbol{\Psi}^{\mathrm{T}}\boldsymbol{\Psi})^{-}\boldsymbol{\Psi}^{\mathrm{T}} - (\boldsymbol{\Psi}^{\mathrm{T}}\boldsymbol{\Psi})^{-}\boldsymbol{\Psi}^{\mathrm{T}}\boldsymbol{\Psi}(\boldsymbol{\Psi}^{\mathrm{T}}\boldsymbol{\Psi})^{-}\boldsymbol{\Psi}^{\mathrm{T}}\}\sigma^2 = \boldsymbol{0}$ (Searle, 1971).

The statistic $SS_E/\sigma^2$ has a $\chi^2$–distribution. From (36), $SS_E$ is a quadratic in $\boldsymbol{y}$. Therefore, if $\boldsymbol{I} - \boldsymbol{\Psi}(\boldsymbol{\Psi}^{\mathrm{T}}\boldsymbol{\Psi})^{-}\boldsymbol{\Psi}^{\mathrm{T}}$ is idempotent, $SS_E/\sigma^2 = \boldsymbol{y}^{\mathrm{T}}\{\boldsymbol{I} - \boldsymbol{\Psi}(\boldsymbol{\Psi}^{\mathrm{T}}\boldsymbol{\Psi})^{-}\boldsymbol{\Psi}^{\mathrm{T}}\}/\sigma^2\,\boldsymbol{y}$ and $var(\boldsymbol{y}) = \boldsymbol{I}\sigma^2$, hence $(1/\sigma^2)\{\boldsymbol{I} - \boldsymbol{\Psi}(\boldsymbol{\Psi}^{\mathrm{T}}\boldsymbol{\Psi})^{-}\boldsymbol{\Psi}^{\mathrm{T}}\}\boldsymbol{I}\sigma^2$ is idempotent as well. So, when $\boldsymbol{y} \sim N(\boldsymbol{\Psi\beta}, \boldsymbol{I}\sigma^2)$, then $\boldsymbol{y}^{\mathrm{T}}\{\boldsymbol{I} - \boldsymbol{\Psi}(\boldsymbol{\Psi}^{\mathrm{T}}\boldsymbol{\Psi})^{-}\boldsymbol{\Psi}^{\mathrm{T}}\}\boldsymbol{y}$ is non-central $\chi^2$ [rank $\{\boldsymbol{I} - \boldsymbol{\Psi}(\boldsymbol{\Psi}^{\mathrm{T}}\boldsymbol{\Psi})^{-}\boldsymbol{\Psi}^{\mathrm{T}}\}$, $(\boldsymbol{\Psi\beta})^{\mathrm{T}}\{\boldsymbol{I} - \boldsymbol{\Psi}(\boldsymbol{\Psi}^{\mathrm{T}}\boldsymbol{\Psi})^{-}\boldsymbol{\Psi}^{\mathrm{T}}\}\boldsymbol{\Psi\beta}/2$], and consequently $SS_E/\sigma^2 \sim \chi^2$ [rank $\{\boldsymbol{I} - \boldsymbol{\Psi}(\boldsymbol{\Psi}^{\mathrm{T}}\boldsymbol{\Psi})^{-}\boldsymbol{\Psi}^{\mathrm{T}}\}$, $\boldsymbol{\beta}^{\mathrm{T}}\boldsymbol{\Psi}^{\mathrm{T}}\{\boldsymbol{I} - \boldsymbol{\Psi}(\boldsymbol{\Psi}^{\mathrm{T}}\boldsymbol{\Psi})^{-}\boldsymbol{\Psi}^{\mathrm{T}}\}\boldsymbol{\Psi\beta}/2\sigma^2$], which, because of properties of a generalized inverse and with rank($\boldsymbol{\Psi}$)=$m$, reduces to

$$SS_E/\sigma^2 \sim \chi^2_{n-m} \quad (46)$$

*3.1. Non-central $\chi^2$'s*

Having shown that $SS_E/\sigma^2$ has a central $\chi^2$–distribution, it can be demonstrated that $SS_R$ has non-central $\chi^2$–distribution. Furthermore, this term is independent of $SS_E$. Thus we are led to $F$-statistic that has non-central $F$-distribution and this in turn is central $F$-distribution under certain null hypotheses.

From (41) $SS_R = \boldsymbol{y}^{\mathrm{T}} \boldsymbol{\Psi} (\boldsymbol{\Psi}^{\mathrm{T}} \boldsymbol{\Psi})^{-} \boldsymbol{\Psi}^{\mathrm{T}} \boldsymbol{y}$ where matrix $\boldsymbol{\Psi} (\boldsymbol{\Psi}^{\mathrm{T}} \boldsymbol{\Psi})^{-} \boldsymbol{\Psi}^{\mathrm{T}}$ is idempotent and its products with matrix $\boldsymbol{I} - \boldsymbol{\Psi} (\boldsymbol{\Psi}^{\mathrm{T}} \boldsymbol{\Psi})^{-} \boldsymbol{\Psi}^{\mathrm{T}}$ are null. Therefore, when $\boldsymbol{y} \sim N(\boldsymbol{\Psi\beta}, \boldsymbol{I}\sigma^2)$, then the quadratic forms $\boldsymbol{y}^{\mathrm{T}} \boldsymbol{\Psi} (\boldsymbol{\Psi}^{\mathrm{T}} \boldsymbol{\Psi})^{-} \boldsymbol{\Psi}^{\mathrm{T}} \boldsymbol{y}$ and $\boldsymbol{y}^{\mathrm{T}} \{\boldsymbol{I} - \boldsymbol{\Psi} (\boldsymbol{\Psi}^{\mathrm{T}} \boldsymbol{\Psi})^{-} \boldsymbol{\Psi}^{\mathrm{T}}\} \boldsymbol{y}$ are distributed independently because of properties of $\boldsymbol{\Psi} (\boldsymbol{\Psi}^{\mathrm{T}} \boldsymbol{\Psi})^{-} \boldsymbol{\Psi}^{\mathrm{T}}$ and $\boldsymbol{\Psi} (\boldsymbol{\Psi}^{\mathrm{T}} \boldsymbol{\Psi})^{-} \boldsymbol{\Psi}^{\mathrm{T}} \boldsymbol{I}\sigma^2 \{\boldsymbol{I} - \boldsymbol{\Psi} (\boldsymbol{\Psi}^{\mathrm{T}} \boldsymbol{\Psi})^{-} \boldsymbol{\Psi}^{\mathrm{T}}\} = \boldsymbol{0}$ (Searle, 1971). So, $SS_R/\sigma^2$ is distributed independently of $SS_E$, with

$$\begin{aligned} SS_R/\sigma^2 &\sim \chi^2 [\mathrm{rank}\,\{\boldsymbol{\Psi} (\boldsymbol{\Psi}^{\mathrm{T}} \boldsymbol{\Psi})^{-} \boldsymbol{\Psi}^{\mathrm{T}}\}, (\boldsymbol{\Psi\beta})^{\mathrm{T}} \boldsymbol{\Psi} (\boldsymbol{\Psi}^{\mathrm{T}} \boldsymbol{\Psi})^{-} \boldsymbol{\Psi}^{\mathrm{T}} \boldsymbol{\Psi\beta}/2\sigma^2] \\ &\sim \chi^2 (m, \boldsymbol{\beta}^{\mathrm{T}} \boldsymbol{\Psi}^{\mathrm{T}} \boldsymbol{\Psi\beta}/2\sigma^2) \end{aligned} \quad (47)$$

Similarly $SS_{Rx}/\sigma^2 = \boldsymbol{y}^{\mathrm{T}} \boldsymbol{X} (\boldsymbol{X}^{\mathrm{T}} \boldsymbol{X})^{-1} \boldsymbol{X}^{\mathrm{T}} \boldsymbol{y}/\sigma^2$, where $\boldsymbol{X} (\boldsymbol{X}^{\mathrm{T}} \boldsymbol{X})^{-1} \boldsymbol{X}^{\mathrm{T}}$ is idempotent and the products of $\boldsymbol{X} (\boldsymbol{X}^{\mathrm{T}} \boldsymbol{X})^{-1} \boldsymbol{X}^{\mathrm{T}}$ and $\boldsymbol{I} - \boldsymbol{\Psi} (\boldsymbol{\Psi}^{\mathrm{T}} \boldsymbol{\Psi})^{-} \boldsymbol{\Psi}^{\mathrm{T}}$ are null since $\boldsymbol{I} - \boldsymbol{\Psi} (\boldsymbol{\Psi}^{\mathrm{T}} \boldsymbol{\Psi})^{-} \boldsymbol{\Psi}^{\mathrm{T}} = \boldsymbol{I} - \boldsymbol{X} (\boldsymbol{X}^{\mathrm{T}} \boldsymbol{X})^{-1} \boldsymbol{X}^{\mathrm{T}} - \boldsymbol{Z} (\boldsymbol{Z}^{\mathrm{T}} \boldsymbol{Z})^{-} \boldsymbol{Z}^{\mathrm{T}}$. Hence $SS_{Rx}$ is distributed independently of $SS_E$ and

$$\begin{aligned} SS_{Rx}/\sigma^2 &\sim \chi^2 [\mathrm{rank}\,\{\boldsymbol{X} (\boldsymbol{X}^{\mathrm{T}} \boldsymbol{X})^{-1} \boldsymbol{X}^{\mathrm{T}}\}, (\boldsymbol{\Psi\beta})^{\mathrm{T}} \boldsymbol{X} (\boldsymbol{X}^{\mathrm{T}} \boldsymbol{X})^{-1} \boldsymbol{X}^{\mathrm{T}} \boldsymbol{\Psi\beta}/2\sigma^2] \\ &\sim \chi^2 \{p+1, \boldsymbol{\beta}^{\mathrm{T}} \boldsymbol{\Psi}^{\mathrm{T}} \boldsymbol{X} (\boldsymbol{X}^{\mathrm{T}} \boldsymbol{X})^{-1} \boldsymbol{X}^{\mathrm{T}} \boldsymbol{\Psi\beta}/2\sigma^2\} \end{aligned} \quad (48)$$

because $\mathrm{rank}\{\boldsymbol{X} (\boldsymbol{X}^{\mathrm{T}} \boldsymbol{X})^{-1} \boldsymbol{X}^{\mathrm{T}}\} = \mathrm{rank}(\boldsymbol{X}) = p+1$.

Analogous argument applies to $SS_{Rc} = \boldsymbol{b}_2^{\mathrm{T}} \boldsymbol{Z}^{\mathrm{T}} \boldsymbol{y} = \boldsymbol{y}^{\mathrm{T}} \boldsymbol{Z} (\boldsymbol{Z}^{\mathrm{T}} \boldsymbol{Z})^{-} \boldsymbol{Z}^{\mathrm{T}} \boldsymbol{y}$, where $\boldsymbol{Z} (\boldsymbol{Z}^{\mathrm{T}} \boldsymbol{Z})^{-} \boldsymbol{Z}^{\mathrm{T}}$ is idempotent and has null products with both $\boldsymbol{X} (\boldsymbol{X}^{\mathrm{T}} \boldsymbol{X})^{-1} \boldsymbol{X}^{\mathrm{T}}$ and $\boldsymbol{I} - \boldsymbol{X} (\boldsymbol{X}^{\mathrm{T}} \boldsymbol{X})^{-1} \boldsymbol{X}^{\mathrm{T}} - \boldsymbol{Z} (\boldsymbol{Z}^{\mathrm{T}} \boldsymbol{Z})^{-} \boldsymbol{Z}^{\mathrm{T}}$. Hence, $SS_{Rc}$ is independent of $SS_{Rx}$ and $SS_E$, and

$$\begin{aligned} SS_{Rc}/\sigma^2 &\sim \chi^2 [\mathrm{rank}\,\{\boldsymbol{Z} (\boldsymbol{Z}^{\mathrm{T}} \boldsymbol{Z})^{-} \boldsymbol{Z}^{\mathrm{T}}\}, (\boldsymbol{\Psi\beta})^{\mathrm{T}} \boldsymbol{Z} (\boldsymbol{Z}^{\mathrm{T}} \boldsymbol{Z})^{-} \boldsymbol{Z}^{\mathrm{T}} \boldsymbol{\Psi\beta}/2\sigma^2] \\ &\sim \chi^2 \{m-p-1, \boldsymbol{\beta}^{\mathrm{T}} \boldsymbol{\Psi}^{\mathrm{T}} \boldsymbol{Z} (\boldsymbol{Z}^{\mathrm{T}} \boldsymbol{Z})^{-} \boldsymbol{Z}^{\mathrm{T}} \boldsymbol{\Psi\beta}/2\sigma^2\} \end{aligned} \quad (49)$$

because $\mathrm{rank}\{\boldsymbol{Z} (\boldsymbol{Z}^{\mathrm{T}} \boldsymbol{Z})^{-} \boldsymbol{Z}^{\mathrm{T}}\} = \mathrm{rank}(\boldsymbol{Z}) = m-p-1$.

*3.2. F-distributions*

Applying the definition of the non-central $F$-distribution to the foregoing results leads to the following $F$-statistics

$$F_R = \frac{SS_R/m}{SS_E/(n-m)} \sim F(m, n-m, \boldsymbol{\beta}^{\mathrm{T}} \boldsymbol{\Psi}^{\mathrm{T}} \boldsymbol{\Psi\beta}/2\sigma^2) \quad (50)$$

$$F_{Rx} = \frac{SS_{Rx}/(p+1)}{SS_E/(n-m)} \sim F\{p+1, n-m, \boldsymbol{\beta}^{\mathrm{T}} \boldsymbol{\Psi}^{\mathrm{T}} \boldsymbol{X} (\boldsymbol{X}^{\mathrm{T}} \boldsymbol{X})^{-1} \boldsymbol{X}^{\mathrm{T}} \boldsymbol{\Psi\beta}/2\sigma^2\} \quad (51)$$

$$F_{Rc} = \frac{SS_{Rc}/(m-p-1)}{SS_E/(n-m)} \sim F\{m-p-1, n-m, \boldsymbol{\beta}^{\mathrm{T}} \boldsymbol{\Psi}^{\mathrm{T}} \boldsymbol{Z} (\boldsymbol{Z}^{\mathrm{T}} \boldsymbol{Z})^{-} \boldsymbol{Z}^{\mathrm{T}} \boldsymbol{\Psi\beta}/2\sigma^2\} \quad (52)$$

Under certain null hypotheses the non-centrality parameters in (50)-(52) are zero, and these non-central $F$'s then become central $F$'s, so providing the statistics for testing those hypotheses.

*3.3. Analysis of variance*

Calculations of the above $F$-statistics can be summarized in analysis of variance tables. For this the calculations for $F_R$ are resumed in Table 2. This table summarizes not only the sums of squares but also the degrees of freedom of the associated $\chi^2$–distributions. In the mean squares, which are sums of squares divided by degrees of freedom, it also shows calculation of the numerator and denominator of $F$. And then the calculation of $F$ itself is shown.

Table 2. Analysis of variance for fitting hybrid data regression.

| Source of Variation | Sum of Squares | Degrees of Freedom | Mean Square | $F$-statistic |
|---|---|---|---|---|
| Regression | $SS_R = \boldsymbol{y}^{\mathrm{T}} \boldsymbol{\Psi}(\boldsymbol{\Psi}^{\mathrm{T}}\boldsymbol{\Psi})^{-}\boldsymbol{\Psi}^{\mathrm{T}}\boldsymbol{y}$ | $m$ | $MS_R = SS_R/m$ | $F_R = MS_R/MS_E$ |
| Residual | $SS_E = \boldsymbol{y}^{\mathrm{T}}\{\boldsymbol{I}-\boldsymbol{\Psi}(\boldsymbol{\Psi}^{\mathrm{T}}\boldsymbol{\Psi})^{-}\boldsymbol{\Psi}^{\mathrm{T}}\}\boldsymbol{y}$ | $n-m$ | $MS_E = SS_E/(n-m)$ | |
| Total | $SS_T = \boldsymbol{y}^{\mathrm{T}}\boldsymbol{y}$ | $n$ | | |

In a manner similar to Table 2 the $F$-ratios of (51) and (52) are summarized in Table 3. And the abbreviated form of this, based on Table 1 and showing only the calculation of (52), is as shown in Table 4.

Table 3. Analysis of variance showing the term of multiple linear regression.

| Source of Variation | Sum of Squares | Degrees of Freedom | Mean Square | $F$-statistic |
|---|---|---|---|---|
| Linear regression | $SS_{Rx} = \boldsymbol{y}^{\mathrm{T}}\boldsymbol{X}(\boldsymbol{X}^{\mathrm{T}}\boldsymbol{X})^{-1}\boldsymbol{X}^{\mathrm{T}}\boldsymbol{y}$ | $p+1$ | $MS_{Rx} = SS_{Rx}/p$ | $F_{Rx} = MS_{Rx}/MS_E$ |
| Corrected Regression | $SS_{Rc} = \boldsymbol{b}_2^{\mathrm{T}}\boldsymbol{Z}^{\mathrm{T}}\boldsymbol{y}$ | $m-p-1$ | $MS_{Rc} = SS_{Rc}/(m-p-1)$ | $F_{Rc} = MS_{Rc}/MS_E$ |
| Residual | $SS_E = \boldsymbol{y}^{\mathrm{T}}\boldsymbol{y}-\boldsymbol{y}^{\mathrm{T}}\boldsymbol{X}(\boldsymbol{X}^{\mathrm{T}}\boldsymbol{X})^{-1}\boldsymbol{X}^{\mathrm{T}}\boldsymbol{y}-\boldsymbol{b}_2^{\mathrm{T}}\boldsymbol{Z}^{\mathrm{T}}\boldsymbol{y}$ | $n-m$ | $MS_E = SS_E/(n-m)$ | |
| Total | $SS_T = \boldsymbol{y}^{\mathrm{T}}\boldsymbol{y}$ | $n$ | | |

Table 4. Analysis of variance corrected for the term of multiple linear regression.

| Source of Variation | Sum of Squares | Degrees of Freedom | Mean Square | $F$-statistic |
|---|---|---|---|---|
| Corrected Regression | $SS_{Rc} = \boldsymbol{b}_2^{\mathrm{T}}\boldsymbol{Z}^{\mathrm{T}}\boldsymbol{y}$ | $m-p-1$ | $MS_{Rc} = SS_{Rc}/(m-p-1)$ | $F_{Rc} = MS_{Rc}/MS_E$ |
| Residual | $SS_E = \boldsymbol{y}^{\mathrm{T}}\boldsymbol{y}-\boldsymbol{y}^{\mathrm{T}}\boldsymbol{X}(\boldsymbol{X}^{\mathrm{T}}\boldsymbol{X})^{-1}\boldsymbol{X}^{\mathrm{T}}\boldsymbol{y}-\boldsymbol{b}_2^{\mathrm{T}}\boldsymbol{Z}^{\mathrm{T}}\boldsymbol{y}$ | $n-m$ | $MS_E = SS_E/(n-m)$ | |
| Corrected Total | $SS_{Tc} = \boldsymbol{y}^{\mathrm{T}}\boldsymbol{y}-\boldsymbol{y}^{\mathrm{T}}\boldsymbol{X}(\boldsymbol{X}^{\mathrm{T}}\boldsymbol{X})^{-1}\boldsymbol{X}^{\mathrm{T}}\boldsymbol{y} = \boldsymbol{\upsilon}^{\mathrm{T}}\boldsymbol{\upsilon}$ | $n-p-1$ | | |

Tables 3 and 4 show the same things for $F_{Rx}$ and $F_{Rc}$ of (51) and (52). But Table 4 presents the abbreviated form of the complete analysis of variance table shown in Table 3. This abbreviated form is derived by removing $SS_{Rx}$ from the body of the table and subtracting it from $SS_T$ to give $SS_{Tc}$, as in Table 1. Thus Table 4 does not contain $F_{Rx}$, but it is identical to Table 3 insofar as $F_{Rc} = MS_{Rc}/MS_E$ is concerned. Although Table 4 is the form in which this analysis of variance may be seen, the most informative is Table 3 which demonstrates how $SS_R$ of Table 2 is partitioned into $SS_{Rx}$ and $SS_{Rc}$, so summarizing both $F_{Rx}$ and $F_{Rc}$.

### *3.4. Pure errors*

In multiple hybrid data regression modelling a model is fitted to data from designed experiment. Here if two or more observations on response variable at the same settings of predictor variables have been completed then a formal test for the lack of fit may be conducted. For this the matrix $\boldsymbol{\Psi}$ should be of full column rank and follow the condition $n>2(p+1)$. If this matrix has $n=2(p+1)$ then the design matrix $\boldsymbol{X}$ from which matrix $\boldsymbol{\Psi}$ is obtained may be augmented by additional set of settings of explanatory variables. For example, the first-order design matrix may be augmented by centre runs which do not impact the usual effect estimates in a $2^k$ design (Myers and Montgomery, 1995).

The lack of fit test requires that we have true replicates on the response $y$ for at least one set of levels on the factors. These replicate points are used to obtain a model-independent estimate of $\sigma^2$. Hence, for $l$ true replicates in which $\boldsymbol{\xi}_0$ and $\boldsymbol{x}_0$ are fixed, the hybrid data model may be written $y_u = f(\boldsymbol{\xi}_0, \boldsymbol{\varphi})\times\phi(\boldsymbol{x}_0, \boldsymbol{\theta})+e_u$, where $u=1, 2,\ldots, l$, implying $\tilde{y} = f(\boldsymbol{\xi}_0, \boldsymbol{\varphi})\times\phi(\boldsymbol{x}_0, \boldsymbol{\theta})+\tilde{e}$, where $\tilde{y} = \sum_{u=1}^{l} y_u / l$ and $\tilde{e} = \sum_{u=1}^{l} e_u / l$. So, whatever theoretical and empirical functions may be, we have $\sum_{u=1}^{l} (y_u - \tilde{y})^2 = \sum_{u=1}^{l} (e_u - \tilde{e})^2$. And if the $e_u$ are independent random variables with variance $\sigma^2$, $E \sum_{u=1}^{l} (y_u - \tilde{y})^2 = E \sum_{u=1}^{l} (e_u - \tilde{e})^2 = (l-1)\,\sigma^2$ (Box and Draper, 1987).

If we have $n$ such sets of replicated runs with $l_i$ runs in the $i^{th}$ set made at $\boldsymbol{\xi}_i$, the individual internal sums of squares may be pooled together to form a pure error sum of squares having as its degrees of freedom the sum of the separate degrees of freedom. Thus the pure error sum of squares is

$$SS_{PE} = \sum_{i=1}^{n} \sum_{u=1}^{l_i} \left(y_{iu} - \tilde{y}_i\right)^2 \tag{53}$$

with degrees of freedom $\sum_{i=1}^{n} \left(l_i - 1\right) = \mathcal{N} - n$, $\mathcal{N} = \sum_{i=1}^{n} l_i$, and an estimator of $\sigma^2$ is $SS_{PE}/(\mathcal{N}-n)$.

In addition, when the hybrid data regression model is fitting to data, the pure error sum of squares is also part of the residual sum of squares. The residual from the $u^{th}$ observation at $\boldsymbol{\xi}_i$ is $y_{iu} - \hat{y}_i = (y_{iu} - \tilde{y}_i) - (\hat{y}_i - \tilde{y}_i)$; on squaring both sides and summing,

$$\sum_{i=1}^{n} \sum_{u=1}^{l_i} \left(y_{iu} - \hat{y}_i\right)^2 = \sum_{i=1}^{n} \sum_{u=1}^{l_i} \left(y_{iu} - \tilde{y}_i\right)^2 + \sum_{i=1}^{n} t_i \left(\hat{y}_i - \tilde{y}_i\right)^2, \tag{54}$$

that is, residual sum of squares = pure error sum of squares + lack of fit sum of squares.

The corresponding equation for degrees of freedom is $\sum_{i=1}^{n} l_i - m = \sum_{i=1}^{n} (l_i - 1) + n - m$. And, we must have $n > m$. If $n=m$, there will be no sum of squares nor degrees of freedom for lack of fit, since the model will always "fit perfectly" (Box and Draper, 1987). However this is feasible for the hybrid data model when the choice of design matrix follows the condition $n \leq 2(p+1)$. For such a model the rank of $\mathcal{N}$ x $2(p+1)$ matrix $\boldsymbol{\Psi}$ is the smallest $m$ for which there exist $\mathcal{N}$ x $m$ matrix $\boldsymbol{\Omega}$ and $m$ x $2(p+1)$ matrix $\boldsymbol{O}$ such that $\boldsymbol{\Psi} = \boldsymbol{\Omega}\,\boldsymbol{O}$. In such full-rank decomposition $m$ is a maximum number of linearly independent rows of $\boldsymbol{\Psi}$ which equals to $n$. Thus for this model there is no lack of fit and the residual sum of squares equals to pure error sum of squares.

A measure of the goodness of fit of the regression is the multiple correlation coefficient, estimated as the product moment correlation between the observed $y_i$'s and the predicted $\hat{y}_i$'s (Searle, 1971). Denoted by $R$, for the hybrid data regression model it can be calculated as

$$R^2 = (\boldsymbol{b}^{\mathrm{T}} \boldsymbol{\Psi}^{\mathrm{T}} \boldsymbol{y} - n\,\bar{y}^2)/(\boldsymbol{y}^{\mathrm{T}} \boldsymbol{y} - n\,\bar{y}^2), \tag{55}$$

where $\bar{y} = n^{-1} \sum_{i=1}^{n} y_i$. This statistic represents the fraction of the variation about the mean that is explained by the fitted models. It is often used as an overall measure of the fit attained (Box and Draper, 1987). No model can explain pure error, so that the maximum possible value of $R^2$ is

$$R^2_{max} = (\boldsymbol{y}^{\mathrm{T}} \boldsymbol{y} - n\,\bar{y}^2 - SS_{PE})/(\boldsymbol{y}^{\mathrm{T}} \boldsymbol{y} - n\,\bar{y}^2). \tag{56}$$

And the hybrid data regression model may have this value for its multiple correlation coefficient because from (41) $\boldsymbol{b}^{\mathrm{T}} \boldsymbol{\Psi}^{\mathrm{T}} \boldsymbol{y} = \boldsymbol{y}^{\mathrm{T}} \boldsymbol{y} - SS_E$ and if the choice of design matrix follows the condition $n \leq 2(p+1)$ then $SS_E = SS_{PE}$, hence from (55) and (56) $R^2 = R^2_{max}$.

### 3.5. Test of hypotheses

For the hybrid data model following equations (50-52) we indicated that those results provide statistics suitable for testing certain hypotheses. As shown in equation (50) the statistic $F_R$ is distributed as a non-central $F$ with non-centrality parameter $\boldsymbol{\beta}^T \boldsymbol{\Psi}^T \boldsymbol{\Psi\beta}/2\sigma^2$. This parameter is zero under the null hypothesis $H_0$: $\boldsymbol{\Psi\beta}=\mathbf{0}$, when $F_R$ then has a central $F$-distribution, $F_{m,\, n-m}$, and can be compared to tabulated values thereof to test that hypothesis. When $F_R \geq$ tabulated $F_{m,\, n-m}$ at the α-level, we reject the null hypothesis $H_0$: $\boldsymbol{\Psi\beta}=\mathbf{0}$, otherwise we do not reject it. Assuming the model $E(\boldsymbol{y}) = \boldsymbol{\Psi\beta}$ we might then say that when $F_R$ is significant the model accounts for a significant portion of the variation in the $y$-variable. But this does not mean that this model, for the particular set of factors, which is the set of explanatory variables of theoretical function, is necessarily the most suitable model: its empirical function may have a subset of those factors which are as significant as the whole, or may include further factors which, when used in combination with some or all of the factors already used, are significantly better than those already used; or there may be the other non-linear functions of these factors that are at least as suitable as the factors as used. None of these contingencies is inconsistent with $F_R$ being significant and the ensuing conclusion that the data are in concordance with the model $E(\boldsymbol{y}) = \boldsymbol{\Psi\beta}$.

Just as $F_R$ provides a test of the model $E(\boldsymbol{y}) = \boldsymbol{\Psi\beta}$, so does $F_{Rc}$ provide a test over and above the multiple linear regression model. In general, therefore, $F_{Rc}$ must be looked on as providing a test of the model $E(\boldsymbol{y}) = \boldsymbol{\Psi\beta}$ over and above the model $E(\boldsymbol{y}) = \boldsymbol{X\theta}$. Since the latter can be considered as fitting the multiple linear regression model to data we look upon $F_{Rc}$ as providing a test of the model $E(\boldsymbol{y}) = \boldsymbol{\Psi\beta}$ over and above the multiple linear regression model. When $F_{Rc}$ is significant we conclude that the model satisfactorily accounts for variation in the $y$-variable. This is not to be taken as evidence that all elements of $\boldsymbol{\beta}$ are non-zero, but only at least one of them, or one linear combination of them, may be. If $F_{Rx}$ has first been found significant, then $F_{Rc}$ being significant indicates that a model with terms in it additional to the multiple linear regression model explains significantly more of the variation in the $y$-variable than does the model $E(\boldsymbol{y})=\boldsymbol{X\theta}$.

The case of both $F_{Rx}$ and $F_{Rc}$ being significant has just been discussed; a further possibility is that $F_{Rx}$ is significant but $F_{Rc}$ is not. This is evidence of the linear regression model being relevant but that fitting the rest of the model does not explain variation in the $y$-variable.

### 3.6. Examining residuals

The vector of residuals

$$\boldsymbol{r}=\boldsymbol{y}-\boldsymbol{\Psi b} \qquad (57)$$

can be plotted in a variety of ways and otherwise investigated to see if assumptions inherent in the assumed model are not being preserved. With matrix $\boldsymbol{I} - \boldsymbol{\Psi}(\boldsymbol{\Psi}^T\boldsymbol{\Psi})^- \boldsymbol{\Psi}^T$, which is symmetric and idempotent, $\boldsymbol{r} = \boldsymbol{y} - \boldsymbol{\Psi b} = \boldsymbol{y} - \boldsymbol{\Psi}(\boldsymbol{\Psi}^T\boldsymbol{\Psi})^- \boldsymbol{\Psi}^T \boldsymbol{y} = \{\boldsymbol{I} - \boldsymbol{\Psi}(\boldsymbol{\Psi}^T\boldsymbol{\Psi})^- \boldsymbol{\Psi}^T\}\boldsymbol{y}$. So, concerning distributional properties of residuals, their expected values are $E(\boldsymbol{r}) = E[\{\boldsymbol{I} - \boldsymbol{\Psi}(\boldsymbol{\Psi}^T\boldsymbol{\Psi})^- \boldsymbol{\Psi}^T\}\, \boldsymbol{y}] = \{\boldsymbol{I} - \boldsymbol{\Psi}(\boldsymbol{\Psi}^T\boldsymbol{\Psi})^- \boldsymbol{\Psi}^T\}\, \boldsymbol{\Psi\beta} = \mathbf{0}$, because $\boldsymbol{\Psi}(\boldsymbol{\Psi}^T\boldsymbol{\Psi})^- \boldsymbol{\Psi}^T\, \boldsymbol{\Psi} = \boldsymbol{\Psi}$, and residual uncertainty matrix is $var(\boldsymbol{r}) = var[\{\boldsymbol{I} - \boldsymbol{\Psi}(\boldsymbol{\Psi}^T\boldsymbol{\Psi})^- \boldsymbol{\Psi}^T\}\boldsymbol{y}] = \{\boldsymbol{I} - \boldsymbol{\Psi}(\boldsymbol{\Psi}^T\boldsymbol{\Psi})^- \boldsymbol{\Psi}^T\}\sigma^2$.

These properties hold true for the residuals of all considered here hybrid data models. Consideration of the extent to which they satisfy other conditions is the means whereby assumptions of the model can be investigated. For example, in assuming normality of the error terms in the model we have $E(\boldsymbol{r}) \sim N\, [\mathbf{0},\, \{\boldsymbol{I} - \boldsymbol{\Psi}(\boldsymbol{\Psi}^T\boldsymbol{\Psi})^- \boldsymbol{\Psi}^T\}\, \sigma^2]$. Plotting the values of $\boldsymbol{r}$ to see if they appear normally distributed therefore provides a means of seeing if the assumption $\boldsymbol{e} \sim N\,(\mathbf{0},\, \boldsymbol{I}\sigma^2)$ might be wrong. In doing this we ignore the fact that because $var(\boldsymbol{r}) = \{\boldsymbol{I} - \boldsymbol{\Psi}(\boldsymbol{\Psi}^T\boldsymbol{\Psi})^- \boldsymbol{\Psi}^T\}\sigma^2$ the $r$'s are correlated since, as Searle (1971) and Anscombe and Tukey (1963) indicate, for at least a two-way table with more than three rows and columns, the effect of correlation among residuals upon graphical procedures is usually negligible.

## 4. Pneumatic gauge modelling

Hybrid data modelling implies concurrent computer simulation and physical experimentation and with this concern let us considers pneumatic gauge example.

Pneumatic gauging is an established technique for determining the location of a quasi-rigid surface (see, for example, Evans and Morgan, 1964; Jackson, 1978; Bridge *et al.*, 2001). The principle of the device is simple. A jet of gauging fluid, typically air, flows from a reservoir at known pressure along a tube fitted with an orifice and out of a convergent nozzle maintained at a known location regarding to a work-piece. The pressure drop between the inlet to the nozzle and the surroundings (i.e., atmosphere) is sensitive to the distance or clearance $S$ between the output tip of the nozzle and the work-piece. The back-pressure $P$ measured by the pressure indicator placed between the orifice and nozzle-work-piece sensor, varies in inverse proportion to the clearance $S$.

Pneumatic gauging involves the determination of dimensional deviations from measurement experiment along with estimates of their associated uncertainties. And in order to extract information from the experimental data, gauge mathematical model is required. The model will need to specify how the gauge is expected to respond to input data and this is the key to extracting information from the data. The quality of information obtained depends directly on the gauge model explanation of its observed behaviour and more dependable uncertainties associated with the quantities of interest.

### *4.1. Gauge theoretical modelling*

The pneumatic gauge can be modelled by algorithmic model which pave the way for application of iterative numerical approximation methods. Continuous modelling may be used to create a representation of steady-flow physical phenomenon at the gauge and therefore gain a better understanding of that phenomenon and enhance the knowledge base of measurement. Gauge theoretical model is a steady-state relationship $P_\lambda=f(S)$ between calculated back-pressure $P_\lambda$ and clearance $S$. This model may be elaborated by considering adiabatic air flows through the both orifice and nozzle-work-piece sensor.

The steady-flow model of the orifice may be derived from the energy equation (Zalmanzon, 1965) and in Mathcad computer code notation (*Mathcad User's Guide*, 1999) is written

$$G_O = C_O B P_S \sqrt{\frac{2g}{RT}} \left| \begin{array}{ll} \sqrt{\dfrac{\gamma}{\gamma-1}\left[\left(\dfrac{P_\lambda}{P_S}\right)^{2/\gamma} - \left(\dfrac{P_\lambda}{P_S}\right)^{(\gamma+1)/\gamma}\right]} & if \quad \dfrac{P_\lambda}{P_S} \ge \left(\dfrac{2}{\gamma+1}\right)^{\gamma/(\gamma-1)} \\ \sqrt{\dfrac{\gamma}{\gamma+1}\left(\dfrac{2}{\gamma+1}\right)^{2/(\gamma-1)}} & otherwise \end{array} \right. , \qquad (58)$$

where $G_O$ is the weight flow rate through the orifice, $B$ is the cross-sectional area of the orifice, which is a simple transformation of its internal diameter $d_O$, namely, $B=\pi\, d_O^2/4$, $g$ is the acceleration of gravity, $\gamma$ is the ratio of specific heats for air, $P_S$ is the pressure in a chamber of compressed air supply, $R$ is the gas constant, $T$ is the absolute temperature, and $C_O$ is the discharge coefficient which takes into account the losses of mechanical energy at the flow entrance area of the orifice and which can be found experimentally only (Zalmanzon, 1965).

The steady-flow model of the nozzle-work-piece sensor is a relationship between response variable - weight flow rate $G$ through the sensor and predictor variables - clearance $S$ and back-pressure $P_\lambda$. Considering the sensor as a short restriction with adiabatic and turbulent flow it may be shown that sensor steady-flow model is derived from the energy equation and in Mathcad computer code notation is written

$$G = CAP_\lambda\sqrt{\frac{2g}{RT}}\left|\begin{array}{ll}\sqrt{\frac{\gamma}{\gamma-1}\left[\left(\frac{P_a}{P_\lambda}\right)^{2/\gamma}-\left(\frac{P_a}{P_\lambda}\right)^{(\gamma+1)/\gamma}\right]} & if \quad \frac{P_a}{P_\lambda}\ge\left(\frac{2}{\gamma+1}\right)^{\gamma/(\gamma-1)} \\ \sqrt{\frac{\gamma}{\gamma+1}\left(\frac{2}{\gamma+1}\right)^{2/(\gamma-1)}} & otherwise\end{array}\right. , \tag{59}$$

where $A$ is the cross-sectional area of the sensor, which is a simple transformation of the clearance $S$, namely, $A=\pi\ d\ S$, $d$ is the inner diameter of nozzle, $P_a$ is the pressure of air in the outlet section of the sensor, and $C$ is the discharge coefficient which takes into account the losses of mechanical energy at the flow entrance area of the sensor.

Gauge theoretical model may be derived by using equations describing adiabatic air flows through the both orifice and nozzle-work-piece sensor. In gauging situation the orifice and nozzle-work-piece sensor are in series. So, for one-dimensional steady flow the weight flow rates through these two restrictors must be equal in order for the total weight of air in the tube between the orifice and nozzle to be constant. Therefore, assuming that the temperature $T$ is constant, the weight flow rate equality through the orifice and nozzle-work-piece sensor in Mathcad notation may be written

$$C_O BP_S\left|\begin{array}{ll}\sqrt{\frac{\gamma}{\gamma-1}\left[\left(\frac{P_\lambda}{P_S}\right)^{2/\gamma}-\left(\frac{P_\lambda}{P_S}\right)^{(\gamma+1)/\gamma}\right]} & if \quad \frac{P_\lambda}{P_S}\ge\left(\frac{2}{\gamma+1}\right)^{\gamma/(\gamma-1)} \\ \sqrt{\frac{\gamma}{\gamma+1}\left(\frac{2}{\gamma+1}\right)^{2/(\gamma-1)}} & otherwise\end{array}\right.$$

$$= CAP_\lambda\left|\begin{array}{ll}\sqrt{\frac{\gamma}{\gamma-1}\left[\left(\frac{P_a}{P_\lambda}\right)^{2/\gamma}-\left(\frac{P_a}{P_\lambda}\right)^{(\gamma+1)/\gamma}\right]} & if \quad \frac{P_a}{P_\lambda}\ge\left(\frac{2}{\gamma+1}\right)^{\gamma/(\gamma-1)} \\ \sqrt{\frac{\gamma}{\gamma+1}\left(\frac{2}{\gamma+1}\right)^{2/(\gamma-1)}} & otherwise\end{array}\right. \tag{60}$$

This theoretical model is given as an implicit algorithmic model in which the variables $P_\lambda$ and $S$ are linked through equation (60). However, by several authors (Zalmanzon 1965, Nakayama 1964, Crnojevic, *et al.*, 1997) it was revealed that the discharge coefficients $C_O$ and $C$ are not constant even for a small variation of such variable quantities as $S$, $P_S$, and $P_\lambda$. So, in this circumstance it is impossible to find least squared estimates of the discharge coefficients and to fit gauge theoretical model to data. Therefore, it should be considered that available gauge physical knowledge is substantially incomplete and, consequently it would be preferable to apply the multiple linear regression modelling. With this approach a multiple linear regression model is used instead of the theoretical model. Such a model does not express any physical knowledge underlying back-pressure variation and reflects only the local influence of explanatory variables on response variable.

### *4.2. Gauge linear regression modelling*

For high precision measurement it is essential the employment of comprehensive, adequate, and precise gauge model. This model is a relationship between observed response variable – measured back-pressure $P$ and explanatory variables; cross-sectional area $A$ of the sensor, absolute pressure $P_S$, and cross-sectional area $B$ of the orifice.

Initially, let us assume that the first-order polynomial model might describe the investigated relationship which is

$$P_1=\theta_0+\theta_1 x_1+\theta_2 x_2+\theta_3 x_3+\varepsilon_1, \quad (61)$$

where $x_1$ represents the coded variable of sensor cross-sectional area $A$, $x_2$ represents the coded variable of absolute pressure $P_S$, $x_3$ represents the coded variable of orifice cross-sectional area $B$, and $\varepsilon_1$ is the error for this model. Also, this polynomial model is a linear function of the unknown parameters $\theta_0$, $\theta_1$, $\theta_2$, $\theta_3$ and the method of least squares is typically used to estimate these regression coefficients (Box and Draper, 1987; Myers and Montgomery, 1995).

The back-pressure $P_1$ is related to three variables that can be controlled in experiment. Table 5 demonstrates the data resulting from an experimental investigation into effect of three variables; cross-sectional area of the sensor, cross-sectional area of the orifice and absolute pressure supply, on gauge back-pressure for the first-order model. Columns 2, 3, and 4 show the levels used for the explanatory variables $A$, $P_S$, and $B$ in natural units of measurements and columns 5, 6 and 7 display the levels in terms of the coded variables $x_1$, $x_2$ and $x_3$.

The least squared estimator of effects $\boldsymbol{\theta}_1$ in the model is $\boldsymbol{q}_1=(\boldsymbol{X}_1^{\,T}\boldsymbol{X}_1)^{-1}\boldsymbol{X}_1^{\,T}\boldsymbol{P}_1$, where $\boldsymbol{X}_1$ is the experimental design $n$ x ($p$+1) matrix of coded factor levels with identity element and $\boldsymbol{P}_1$ is the $n$ x 1 vector of back-pressure observations. The estimate of effects $\boldsymbol{\theta}_1$ is $\boldsymbol{q}_1$= (208.423 –34.409 36.616 18.277)$^T$. So, the least squares fit of first-order regression model is

$$P_{1fit}=208.423-34.409x_1+36.616x_2+18.277x_3 \quad (62)$$

The $F$-test for significance of regression is a test to determine if there is a linear relationship between the back-pressure and a set of coded variables $x_1$, $x_2$, and $x_3$. The test is summarized in Table 6. And, because $\alpha$=0.05 has been selected, then we have $F_0=17.85>F_{0.05,\ 3,\ 7}=4.35$. So, we would reject the hypothesis $H_0$: $\theta_1=\theta_2=\theta_3=0$, concluding that at least some of these parameters are nonzero.

Table 5. Two-level factorial design and data for pneumatic gauge modelling

| Run | $A$, mm$^2$ | $P_S$, MPa | $B$, mm$^2$ | $x_1$ | $x_2$ | $x_3$ | $\boldsymbol{P}_1$, kPa | $\boldsymbol{P}_{1fit}$, kPa | $\boldsymbol{P}_\lambda$, kPa | $\boldsymbol{P}_{\lambda fit}$, kPa | $\boldsymbol{P}_\nu$, kPa | $\boldsymbol{P}_{\nu fit}$, kPa |
|---|---|---|---|---|---|---|---|---|---|---|---|---|
| 1 | 0.251 | 0.199 | 0.503 | -1 | -1 | -1 | 189.487 | 187.939 | 187.986 | 188.345 | 187.410 | 188.704 |
| 2 | 1.257 | 0.199 | 0.503 | 1 | -1 | -1 | 126.332 | 119.121 | 115.955 | 126.913 | 116.513 | 126.729 |
| 3 | 0.251 | 0.297 | 0.503 | -1 | 1 | -1 | 283.336 | 261.170 | 280.554 | 283.472 | 279.595 | 283.427 |
| 4 | 1.257 | 0.297 | 0.503 | 1 | 1 | -1 | 155.164 | 192.352 | 134.781 | 154.861 | 136.175 | 154.948 |
| 5 | 0.251 | 0.199 | 1.131 | -1 | -1 | 1 | 197.724 | 224.493 | 196.727 | 198.295 | 196.582 | 198.099 |
| 6 | 1.257 | 0.199 | 1.131 | 1 | -1 | 1 | 167.422 | 155.675 | 155.951 | 166.684 | 155.431 | 166.922 |
| 7 | 0.251 | 0.297 | 1.131 | -1 | 1 | 1 | 294.516 | 297.724 | 293.607 | 294.225 | 293.388 | 294.322 |
| 8 | 1.257 | 0.297 | 1.131 | 1 | 1 | 1 | 240.874 | 228.906 | 229.213 | 240.605 | 226.924 | 240.681 |
| 9 | 0.754 | 0.248 | 0.817 | 0 | 0 | 0 | 213.317 | 208.423 | 206.223 | 213.083 | 204.463 | 212.938 |
| 10 | 0.754 | 0.248 | 0.817 | 0 | 0 | 0 | 212.532 | 208.423 | 206.223 | 213.083 | 204.463 | 212.938 |
| 11 | 0.754 | 0.248 | 0.817 | 0 | 0 | 0 | 211.944 | 208.423 | 206.223 | 213.083 | 204.463 | 212.938 |

Table 6. Analysis of variance for the first-order regression model

| Source of Variation | Sum of Squares, (kPa)$^2$ | Degrees of Freedom | Mean Square, (kPa)$^2$ | $F_0$ |
|---|---|---|---|---|
| Regression | $SSr$=2.287x10$^4$ | 3 | MSr=7.623x10$^3$ | 17.85 |
| Error | $SSe$=2.99x10$^3$ | 7 | MSe=427.148 | |
| Lack of Fit | $SS_{LoF}$=2.989x10$^3$ | 5 | $MS_{LoF}$=597.818 | 1260 |
| Pure Error | $SS_{PE}$=0.949 | 2 | $MS_{PE}$=0.474 | |
| Total | $Syy$=2.586x10$^4$ | 14 | | |

As shown in Table 5, the experimental design has three replicated centre runs. So, the residual sum of squares can be partitioned into pure error and lack of fit components. The lack-of-fit test in Table 6 tests the lack of fit for the first-order regression model. The *F*-value for this test is large ($F_0$=1260>$F_{0.05,\ 5,\ 2}$=5.79), implying that this regression model is inadequate.

In terms of equation (13) for adequacy attainment the simpler first-order model with function $\boldsymbol{X}_1\boldsymbol{\theta}_1$ of first-order may be added by the $\boldsymbol{X}_2\boldsymbol{\theta}_2$ terms of second-order which perhaps would represent the response adequately. So, it could be assumed that the second-order polynomial model might describe the investigated relationship which is

$$P_2=\theta_0+\theta_1x_1+\theta_2x_2+\theta_3x_3+\theta_{11}x_1^2+\theta_{22}x_2^2+\theta_{33}x_3^2+\theta_{12}x_1x_2+\theta_{13}x_1x_3+\theta_{23}x_2x_3+\varepsilon_2 \quad (63)$$

Table 7 demonstrates the data resulting from an empirical investigation into effect of the three variables on gauge back-pressure $P_2$ for the second-order model. The least squared estimator of effects $\boldsymbol{\theta}$ in the second-order model is $\boldsymbol{q}=(\boldsymbol{X}_2^{\ T}\boldsymbol{X}_2)^{-1}\boldsymbol{X}_2^{\ T}\boldsymbol{P}_2$, where $\boldsymbol{X}_2$ is the Box-Behnken design matrix of coded factor levels with identity element and $\boldsymbol{P}_2$ is the *n*x1 vector of back-pressure observations. The estimate of effects $\boldsymbol{\theta}$ is $\boldsymbol{q}$= (212.598 –34.274 38.221 21.697 0.286 –2.362 –6.333 –9.561 13.288 6.227) $^{T}$. So, the least squares fit of the second-order regression model is

$$P_{2fit}=212.598–34.274x_1+38.221x_2+21.697x_2+0.286x_1^2–2.362x_2^2–6.333x_2^2 \\ –9.561x_1x_2+13.288x_1x_3+6.227x_2x_3 \quad (64)$$

The *F*-test for significance of regression is summarized in Table 8 and, because α=0.05 has been selected, then $F_0$=118.419>$F_{0.05,\ 9,\ 5}$ =4.77. So, we would reject the hypothesis $H_0$: $\theta_1=\theta_2=\theta_3=\theta_{11}=\theta_{22}=\theta_{33}=\theta_{12}=\theta_{13}=\theta_{23}=0$, concluding that at least some of these parameters are nonzero.

Table 7. Box-Behnken design and data for the second-order regression model

| Run | *A*, $mm^2$ | $P_S$, MPa | *B*, $mm^2$ | $x_1$ | $x_2$ | $x_3$ | $\boldsymbol{P}_2$, kPa | $\boldsymbol{P}_{2fit}$, kPa |
|---|---|---|---|---|---|---|---|---|
| 1 | 0.251 | 0.199 | 0.817 | -1 | -1 | 0 | 195.273 | 195.273 |
| 2 | 0.251 | 0.297 | 0.817 | -1 | 1 | 0 | 292.260 | 292.260 |
| 3 | 0.546 | 0.199 | 0.817 | 1 | -1 | 0 | 147.907 | 147.907 |
| 4 | 1.834 | 0.297 | 0.817 | 1 | 1 | 0 | 206.648 | 206.648 |
| 5 | 0.251 | 0.248 | 0.503 | -1 | 0 | -1 | 237.147 | 237.147 |
| 6 | 0.251 | 0.248 | 1.131 | -1 | 0 | 1 | 246.561 | 246.561 |
| 7 | 1.257 | 0.248 | 0.503 | 1 | 0 | -1 | 139.963 | 139.963 |
| 8 | 1.257 | 0.248 | 1.131 | 1 | 0 | 1 | 202.530 | 202.530 |
| 9 | 0.754 | 0.199 | 0.503 | 0 | -1 | -1 | 147.220 | 147.220 |
| 10 | 0.754 | 0.199 | 1.131 | 0 | -1 | 1 | 185.564 | 185.564 |
| 11 | 0.754 | 0.297 | 0.503 | 0 | 1 | -1 | 209.787 | 209.787 |
| 12 | 0.754 | 0.297 | 1.131 | 0 | 1 | 1 | 273.039 | 273.039 |
| 13 | 0.754 | 0.248 | 0.817 | 0 | 0 | 0 | 213.317 | 212.565 |
| 14 | 0.754 | 0.248 | 0.817 | 0 | 0 | 0 | 212.532 | 212.565 |
| 15 | 0.754 | 0.248 | 0.817 | 0 | 0 | 0 | 211.944 | 212.565 |

Table 8. Analysis of variance for gauge second-order regression model

| Source of Variation | Sum of Squares, $(kPa)^2$ | Degrees of Freedom | Mean Square, $(kPa)^2$ | $F_0$ |
|---|---|---|---|---|
| Regression | $SSr$=2.624x$10^4$ | 9 | $MSr$=2.916x$10^3$ | 118.419 |
| Error | $SSe$=123.114 | 5 | $MSe$=24.623 | |
| Lack of Fit | $SS_{LoF}$=122.165 | 3 | $MS_{LoF}$=40.722 | 85.831 |
| Pure Error | $SS_{PE}$=0.949 | 2 | $MS_{PE}$=0.474 | |
| Total | $Syy$=2.637x$10^4$ | 14 | | |

As shown in Table 7, the experimental design has three replicated centre runs. So, the residual sum of squares can be partitioned into pure error and lack of fit components. The lack-of-fit test in Table 8 tests the lack of fit for quadratic model. The *F*-value for this test is large ($F_0$=85.831>$F_{0.05,\ 3,\ 2}$=19.16), implying that the second-order regression model is inadequate as well.

Numerical model validation indicates that the residual sum of squares which remains unexplained by this model is 123.114(kPa)$^2$ and the sample standard deviation is 2.965kPa. Hence, this confirms that even with second-order linear regression model only the low level of precision may be attained. This regression model has low precision because implies big standard deviation which indicates that structural gauge model should be improved in some way.

### *4.3. Gauge hybrid data modelling*

From above consideration follows that in terms of model (13) the simpler first-order model with function $\boldsymbol{X}_1\boldsymbol{\theta}_1$ was added by the $\boldsymbol{X}_2\boldsymbol{\theta}_2$ terms of second-order but it has not represented the response adequately. So, for model adequacy attainment the other option has been entertained by adding the terms of $\boldsymbol{Y}_1\boldsymbol{\theta}_1$ instead of $\boldsymbol{X}_2\boldsymbol{\theta}_2$. Applying computer-simulation and empirical data for model building and solving it has been accepted that the theoretical adiabatic model tolerably presents the available physical knowledge. However, knowing that the second-order linear regression model does not fit adequately the empirical data, the exact unknown model has been approximated by the product of polynomial empirical function and given theoretical function (60). The theoretical implicit function may be represented explicitly by $P_\lambda=f(A, P_S, B; \boldsymbol{\lambda})$, where the vector $\boldsymbol{\lambda}=(\gamma\ \ C\ \ C_0)^{\mathrm{T}}$. So, gauge hybrid data model may be written

$$P_1 = f(A, P_S, B; \boldsymbol{\lambda}) \times \phi(\boldsymbol{x}, \boldsymbol{\theta}_{1\lambda})+e_\lambda, \qquad (65)$$

where $\boldsymbol{x}$ is the vector of coded variables $x_1$, $x_2$, $x_3$ and the function $\phi(\boldsymbol{x}, \boldsymbol{\theta}_{1\lambda})$ is specified as the first-order polynomial which in coded variables is written

$$\phi(x_1, x_2, x_3; \boldsymbol{\theta}_{1\lambda}) = \theta_{0\lambda}+\theta_{1\lambda}x_1+\theta_{2\lambda}x_2+\theta_{3\lambda}x_3 \qquad (66)$$

For gauge hybrid data model building the physical designed experiment shown in Table 5 has been employed. So, here the first-order experimental design with corresponding levels of explanatory variables was used for hybrid data modelling and to ensure the design of full column rank it was supplied with three replicated centre runs. The same set of levels of explanatory variables was used for the corresponding deterministic computer-simulation experiment which has been carried out using the theoretical function (60). In this experiment we were interested in obtaining theoretical data, so the values of discharge coefficients have been accepted as $C_0$=$C$=1, yielding the response vector $\boldsymbol{P}_\lambda$ presented in Table 5. Then, vector $\boldsymbol{P}_\lambda$ has been transformed into diagonal matrix $\boldsymbol{D}_\lambda=\mathrm{diag}(\boldsymbol{P}_\lambda)$.

Gauge hybrid data model has been rewritten as $\boldsymbol{P}_1= \boldsymbol{X\theta}_1 +\boldsymbol{Y}_\lambda\boldsymbol{\theta}_{1\lambda} +\boldsymbol{e}_\lambda$, where $\boldsymbol{X}$ is the two-level factorial design matrix with identity element and $\boldsymbol{Y}_\lambda= (\boldsymbol{D}_\lambda-\boldsymbol{I})\ \boldsymbol{X}$. Next the model was transformed into equation $\boldsymbol{P}_1= \boldsymbol{\Psi}_\lambda\boldsymbol{\beta}_\lambda +\boldsymbol{e}_\lambda$, where $\boldsymbol{\Psi}_\lambda=[\boldsymbol{X}\ \boldsymbol{Y}_\lambda]$ and $\boldsymbol{\beta}_\lambda=\begin{bmatrix}\boldsymbol{\theta}_1\\ \boldsymbol{\theta}_{1\lambda}\end{bmatrix}$, and its least squared solution was obtained using equation $\boldsymbol{b}_\lambda= (\boldsymbol{\Psi}_\lambda^{\mathrm{T}}\boldsymbol{\Psi}_\lambda)^{-1}\boldsymbol{\Psi}_\lambda^{\mathrm{T}}\boldsymbol{P}_1$. The result is $\boldsymbol{b}_\lambda$= (27.044 4.607 6.614 3.894 0.907 −0.012 −0.010 −0.016)$^{\mathrm{T}}$. Here the matrix $\boldsymbol{X}$ was selected so that the matrix $\boldsymbol{\Psi}_\lambda^{\mathrm{T}}\boldsymbol{\Psi}_\lambda$ has an inverse and, hence the vector $\boldsymbol{b}_\lambda$ is an estimate of effects $\boldsymbol{\beta}_\lambda$.

For model validation the computation of fitted hybrid data model $\boldsymbol{P}_{\lambda fit}=\boldsymbol{\Psi}_\lambda\boldsymbol{b}_\lambda$ was completed and the obtained result shown in Table 5. Residual analysis was used for model validation and the residuals from the fitted model were computed by equation $\boldsymbol{R}_\lambda=\boldsymbol{P}_1-\boldsymbol{\Psi}_\lambda\boldsymbol{b}_\lambda$. The normal plot of residuals for gauge hybrid data model is shown in Figure 1 and the points at the plot form approximately straight line. So, we can accept the residuals are normally distributed and apply the elaborated hybrid data regression analysis.

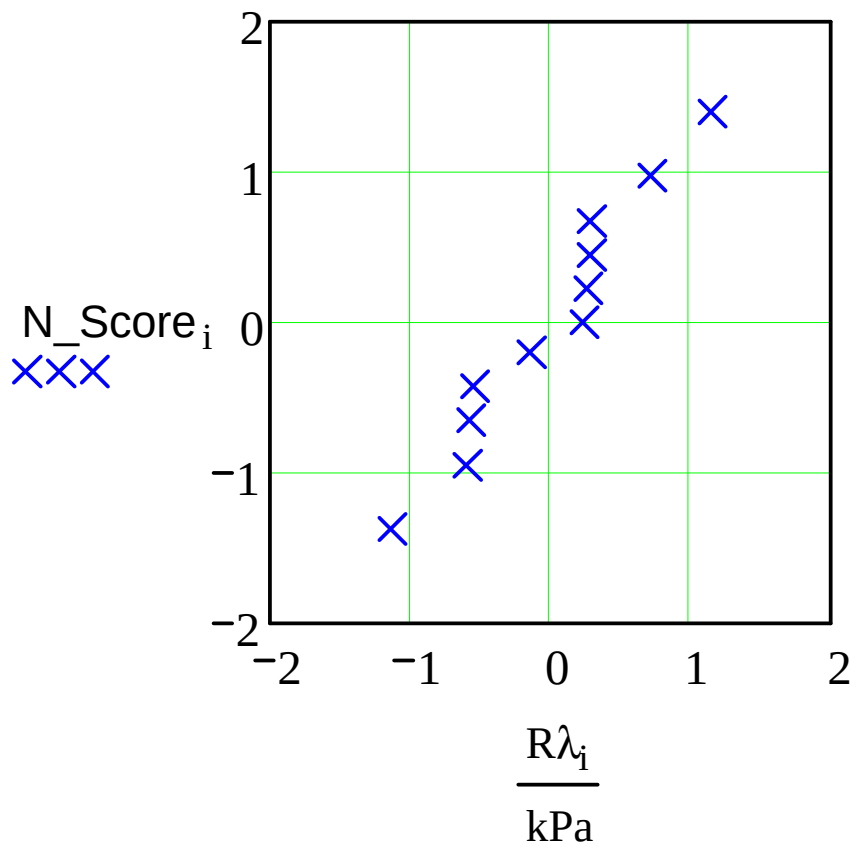

Figure 1. Normal probability plot of residuals for gauge adiabatic hybrid data model

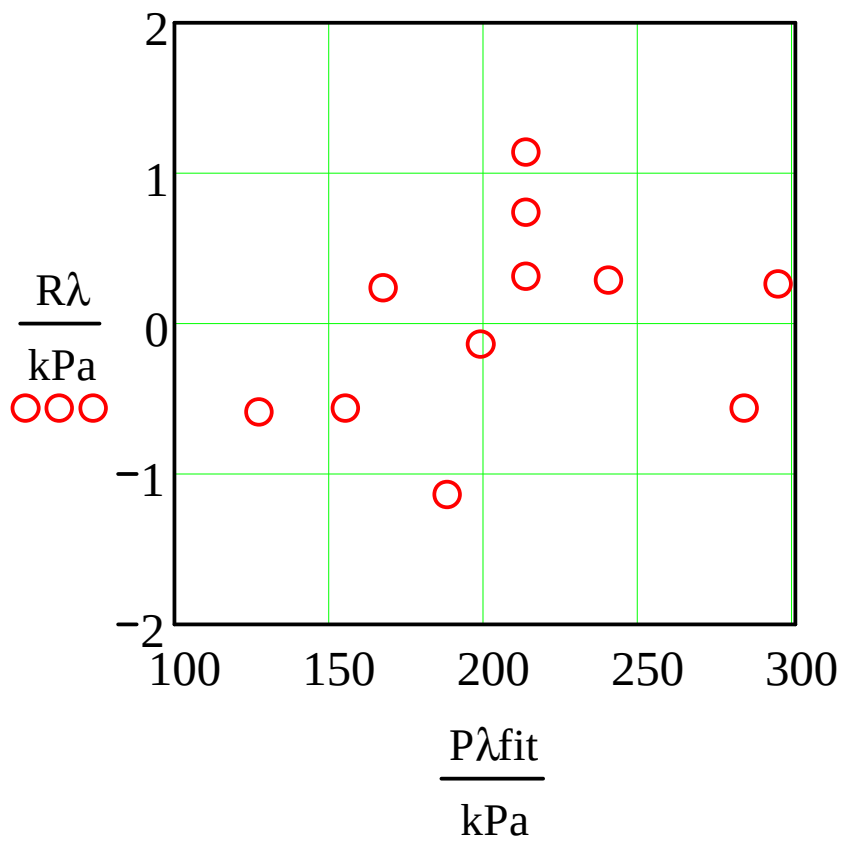

Figure 2. Plot of residuals versus response predicted values

The scatter plot of residuals versus predicted values is the primary plot which may be used to assess sufficiency of the functional part of the model (NIST/SEMATECH *e-Handbook of Statistical Methods*, 2006). The plot shown in Figure 2 demonstrates the residuals approximate the random errors that make the relationship between predictor and response variables a statistical relationship and suggest that main effects hybrid data model fits the data.

Using equations presented in Table 3 the $F$-test for significance of regression has been summarized in Table 9 and, because $\alpha$=0.05 was selected, then $F_{Rx}$=84730>>$F_{0.05,\ 4,\ 3}$=9.12 and $F_{Rc}$ =505> $F_{0.05,\ 4,\ 3}$=9.12. Here $F_{Rx}$ has first been found significant and then $F_{Rc}$ being significant indicate that the model with terms in it additional to the first-order linear regression model explains significantly more of the variation in the $P_1$-variable than the model $E(\boldsymbol{P}_2)$ =$\boldsymbol{X\theta}$. Also, the lack-of-fit test in Table 9 tests the lack of fit for gauge main effects hybrid data model. The $F$-value for this test ($F_{LoF}$=7.342<$F_{0.05,\ 1,\ 2}$ =18.51) is small, implying that the main effects hybrid data model is adequate.

Table 9. Analysis of variance for main effects adiabatic hybrid data model.

| Source of Variation | Sum of Squares, $(kPa)^2$ | Degree of Freedom | Mean Square, $(kPa)^2$ | $F$ |
|---|---|---|---|---|
| Regression x | $SS_{Rx}$ =5.007x$10^5$ | 4 | $MS_{Rx}$ =1.252x$10^5$ | 84730 |
| Regression c | $SS_{Rc}$ = 2986 | 4 | $MS_{Rc}$ = 746.402 | 505 |
| Residual | $SS_E$ = 4.432 | 3 | $MS_E$ = 1.477 | |
| Lack of Fit | $SS_{LoF}$ = 3.483 | 1 | $MS_{LoF}$ = 3.483 | 7.342 |
| Pure Error | $SS_{PE}$ = 0.949 | 2 | $MS_{PE}$ = 0.474 | |
| Total | $SS_T$ = 5.037x$10^5$ | 11 | | |

However, earlier some analytical work has been done on developing criteria for judging the adequacy of regression model from a prediction point of view. The Box and Wetz (1973) work suggests that the observed $F$-ratio must be at least four or five times the critical value from the $F$-table if the regression model is to be useful as a predictor (Myers and Montgomery, 1995). From above for adiabatic hybrid data model the observed $F$-ration is 2.5 times the critical value which is not follow to the Box and Wetz suggestion. Therefore, the way of hybrid data model improvement as a predictor may be connected with the other theoretical function employment.

The steady-flow model (58) of the orifice was derived from the energy equation assuming that the air-flow process is adiabatic. But if isochoric air-flow process is presumed then in Mathcad computer code notation the steady-flow model for the orifice is written

$$G_{O\nu} = C_O B\sqrt{\frac{2g}{RT}}\left|\begin{array}{l}\sqrt{P_\nu\left(P_S - P_\nu\right)} \quad if \quad \frac{P_\nu}{P_S} \geq 0.5 \\ \frac{P_S}{2} \quad otherwise\end{array}\right. \tag{67}$$

and the isochoric steady-flow model of the nozzle-work-piece sensor is

$$G_\nu = CA\sqrt{\frac{2g}{RT}}\left|\begin{array}{l}\sqrt{P_a\left(P_\nu - P_a\right)} \quad if \quad \frac{P_a}{P_\nu} \geq 0.5 \\ \frac{P_\nu}{2} \quad otherwise\end{array}\right. , \tag{68}$$

where $G_{O\nu}$ and $G_\nu$ are the weight flow rates through the orifice and sensor, respectively, and $P_\nu$ is the calculated back-pressure for isochoric air-flow.

In this case the steady-state model of the gauge may be derived by using equations describing the isochoric air flows through the both orifice and nozzle-work-piece sensor. And for one-dimensional steady flow the weight flow rates through these two restrictors must be equal in order for the total weight of air in the tube between the orifice and nozzle to be constant. Therefore, in this case the weight flow rate equality is written

$$C_O B\left|\begin{array}{l}\sqrt{P_\nu\left(P_S - P_\nu\right)} \quad if \quad \frac{P_\nu}{P_S} \geq 0.5 \\ \frac{P_S}{2} \quad otherwise\end{array}\right. = CA\left|\begin{array}{l}\sqrt{P_a\left(P_\nu - P_a\right)} \quad if \quad \frac{P_a}{P_\nu} \geq 0.5 \\ \frac{P_\nu}{2} \quad otherwise\end{array}\right. \tag{69}$$

This implicit algorithmic model may be represented explicitly by equation $P_\nu = f(A, P_S, B; \boldsymbol{\nu})$, where the vector $\boldsymbol{\nu} = (C \quad C_0)^T$. So, gauge isochoric hybrid data model may be written

$$P_1 = f(A, P_S, B; \boldsymbol{\nu}) \times \phi(\boldsymbol{x}, \boldsymbol{\theta}_{1\nu}) + e_\nu \tag{70}$$

Here for hybrid data model building the physical experiment shown in Table 5 has been employed as well. And in parallel to the physical experiment the corresponding deterministic computer-simulation experiment has been carried out using algorithmic function (69) with $C_0 = C = 1$ and yielding the response vector $\boldsymbol{P}_\nu$ presented in Table 5. Further, employing the matrix $\boldsymbol{D}_\nu = \mathrm{diag}(\boldsymbol{P}_\nu)$, gauge hybrid data model has been rewritten as $\boldsymbol{P}_1 = \boldsymbol{X\theta}_1 + \boldsymbol{Y}_\nu\boldsymbol{\theta}_{1\nu} + \boldsymbol{e}_\nu$, where $\boldsymbol{Y}_\nu = (\boldsymbol{D}_\nu - \boldsymbol{I})\boldsymbol{X}$. After that the model has been transformed into equation $\boldsymbol{P}_1 = \boldsymbol{\Psi}_\nu\boldsymbol{\beta}_\nu + \boldsymbol{e}_\nu$, where $\boldsymbol{\Psi}_\nu = [\boldsymbol{X}\ \boldsymbol{Y}_\nu]$ and $\boldsymbol{\beta}_\nu = \begin{bmatrix}\boldsymbol{\theta}_1 \\ \boldsymbol{\theta}_{1\nu}\end{bmatrix}$, and its solution by equation $\boldsymbol{b}_\nu = (\boldsymbol{\Psi}_\nu^T\boldsymbol{\Psi}_\nu)^{-1}\boldsymbol{\Psi}_\nu^T\boldsymbol{P}_1$ was $\boldsymbol{b}_\nu =$ $(15.429 \quad 5.647 \quad 7.694 \quad 2.555 \quad 0.971 \ -0.006 \ -0.026 \ -0.013)^T$.

For model validation the computation of fitted hybrid data model $\boldsymbol{P}_{\nu fit} = \boldsymbol{\Psi}_\nu\boldsymbol{b}_\nu$ was completed and obtained result shown in Table 5 by vector $\boldsymbol{P}_{\nu fit}$. Residual analysis was used for model validation and the residuals from the fitted model were computed by equation $\boldsymbol{R}_\nu = \boldsymbol{P}_1 - \boldsymbol{\Psi}_\nu\boldsymbol{b}_\nu$. The normal plot of residuals for gauge isochoric hybrid data model is shown in Figure 3 and the points at the plot form approximately straight line. So, we can admit the residuals are normally distributed and apply the hybrid data regression analysis.

The scatter plot of residuals versus predicted values shown in Figure 4 demonstrates the residuals approximate the random errors that make the relationship between predictor and response variables a statistical relationship and suggest that main effects hybrid data model fits the data.

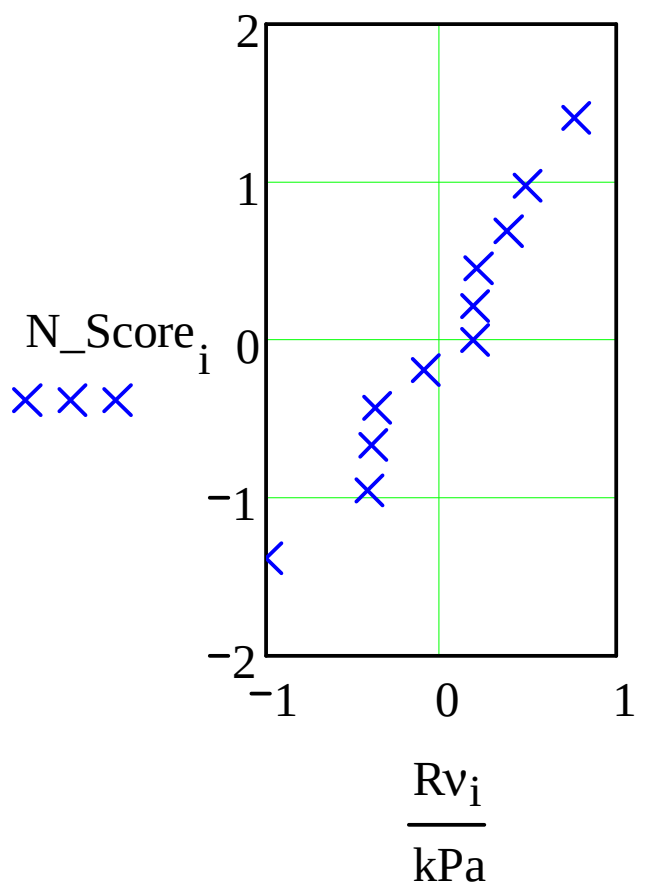

Figure 3. Normal probability plot of residuals for gauge isochoric hybrid data model

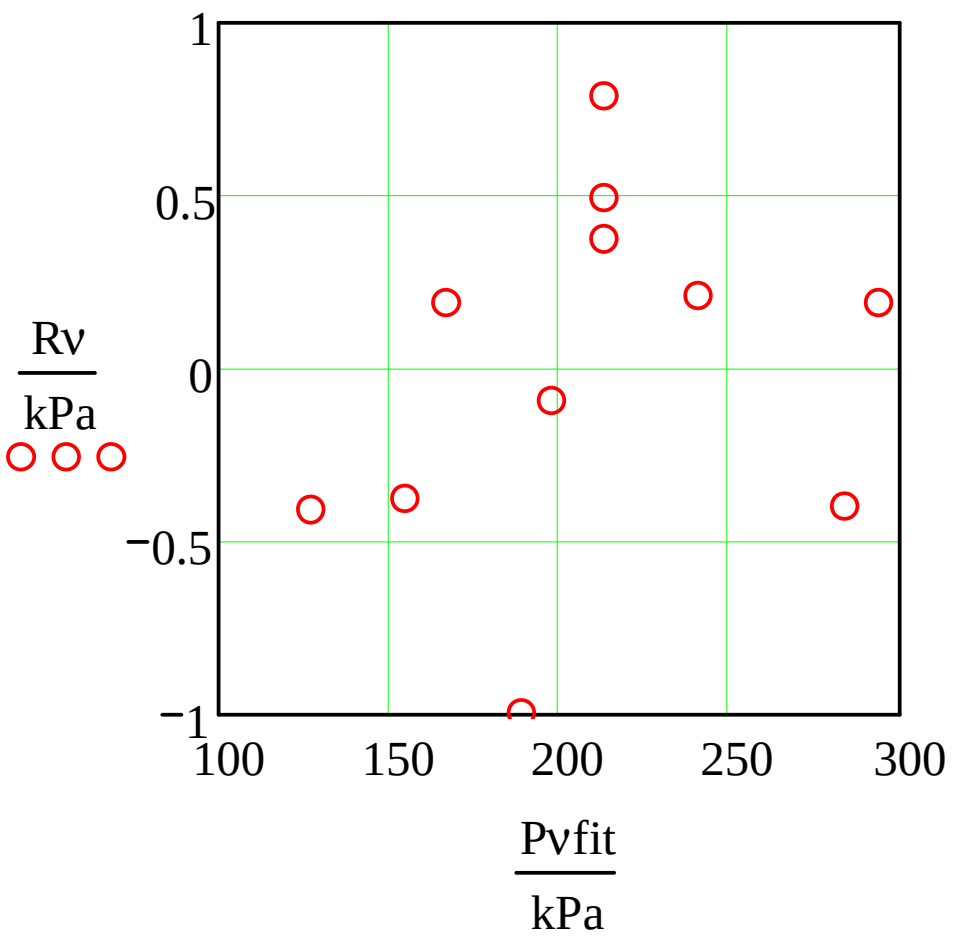

Figure 4. Plot of residuals versus response predicted values

The $F$-test for significance of regression has been summarized in Table 10 and, because $\alpha$=0.05 was selected, then $F_{Rx}$=145200>>$F_{0.05,\ 4,\ 3}$=9.12 and $F_{Rc}$ =866> $F_{0.05,\ 4,\ 3}$=9.12. Here $F_{Rx}$ has first been obtained significant and then $F_{Rc}$ being significant indicate that the model with terms in it additional to the first-order linear regression model explains significantly more of the variation in the $P_1$-variable than the model $E(\boldsymbol{P}_2) = \boldsymbol{X\theta}$. Also, the lack-of-fit test in Table 10 tests the lack of fit for gauge main effects hybrid data model. The $F$-value for this test is small ($F_{LoF}$=3.45<$F_{0.05,\ 1,\ 2}$ =18.51), implying that the hybrid data model is adequate.

Table 10. Analysis of variance for main effects isochoric hybrid data model.

| Source of Variation | Sum of Squares, $(kPa)^2$ | Degree of Freedom | Mean Square, $(kPa)^2$ | $F$ |
|---|---|---|---|---|
| Regression x | $SS_{Rx}$ =5.007x$10^5$ | 4 | $MS_{Rx}$ =1.252x$10^5$ | 145200 |
| Regression c | $SS_{Rc}$ = 2987 | 4 | $MS_{Rc}$ = 746.863 | 866 |
| Residual | $SS_E$ = 2.586 | 3 | $MS_E$ = 0.862 | |
| Lack of Fit | $SS_{LoF}$ = 1.637 | 1 | $MS_{LoF}$ = 1.637 | 3.45 |
| Pure Error | $SS_{PE}$ = 0.949 | 2 | $MS_{PE}$ = 0.474 | |
| Total | $SS_T$ = 5.037x$10^5$ | 11 | | |

Here for isochoric hybrid data model the observed $F$-ration is 5.4 times the critical value which is follow to the Box and Wetz suggestion. Also, this hybrid data model encapsulates the available physical gauge knowledge. For this adequate model the unexplained residual sum of squares remains only 2.586$(kPa)^2$ which is 47.6 times less than for the second-order multiple linear regression model and can be considered as an indicated quantity for superior model quality. Furthermore, the sample standard deviation here is only 0.509kPa, which is 5.8 times less than for gauge multiple linear regression model. Hence, the main effects isochoric hybrid data model has the best precision because implies the smallest standard deviation and emulates the gauge with minimum uncertainty.

## 5. Discussion and conclusions

Undoubtedly the subject of statistical model construction or identification is heavily dependent on the results of theoretical analysis of the object under observation (Akaike, 1974). And since all models are wrong it is impossible to obtain a "correct" one by excessive elaboration (Box, 1976). On the contrary following William of Occam the hybrid data modelling can represent an economical description of natural phenomena.

In iterative physical experimentation usually the initial design is run and then, in the light of the results obtained, it is decided what should be done next. But whatever is done must be done by using the expert subject-matter knowledge of the investigator whose input can contribute to possible changes in the model, in the input variables, in the measured responses and in the objective (Box, 1994). Therefore, in concurrent physical and computer experimentation the initial design needs to be run physically and computationally and then, using statistical inference it is decided what should be done next. Here, for adequacy attainment, instead of empirical component complication the theoretical and simple empirical components may be combined into hybrid data regression model. By means of hybrid data regression modelling it is possible to incorporate quantitatively the available subject-matter knowledge into measurement system regression model. This new type of model building and solving techniques overcomes the main drawback of multiple linear regression modelling, connected with the lack of theoretical information and data about underlying mechanisms, and let us a possibility to elaborate adequate, precise, and parsimonious models which can be used for better measurement system understanding and its performance enhancement. Hybrid data multiple regression modelling is a procedure for data fusion from computer-simulation and physical designed experiments with the aim of maximizing the measurement information content.

The theory of inference regarding parameter estimation generally assumes that the true model for a given set of physical data is known and pre-specified. In practice a model may be formulated from these data, and it is increasingly common for tens or even hundreds of possible models to be entertained (data mining) (Chatfield, 1995). But even when a model is pre-specified on subject-matter grounds and then corrected employing empirical data, it may be formulated incorrectly. Thus, incorporating data from computer experiment let us a possibility to elaborate parsimonious and adequate models with the use of less empirical data. Finally, a single hybrid data model which entertains statistically significant subject-matter knowledge and data may be selected as a ‘winner’ when the other models give inadequate or unsatisfactory fit.

In hybrid data modelling the statistical inference is broadened to include theoretical model formulation, but it is not clear to what extent we can formalize the steps taken by an analyst during hybrid data generation and model building. However, the hybrid data modelling adapts naturally to cope with subject-matter knowledge uncertainty. This type of modelling gives an opportunity to start working with such uncertainty and to give due regard to the computer-based revolution in model formulation which has taken place. With hybrid data multiple regression modelling we have an advantage to extend the formal statistical inference beyond standard regression techniques to theoretical considerations. And, by incorporating statistically significant computer-simulation data it is possible to improve model performance significantly.

The elaborated techniques have been successfully employed for adiabatic and isochoric hybrid data modelling of the pneumatic gauge. The application of these techniques, using the numerical correctness of Mathcad software package, has revealed their effectiveness for elaboration of parsimonious and adequate models. The present investigation has elucidated that if the modelling is accomplished employing concurrently the two sources of data from computer-simulation and physical designed experiments; it is possible to achieve significant model improvement. The validation of gauge adiabatic and isochoric hybrid data multiple regression models has confirmed that these models fit the reference empirical data much better than the second-order multiple linear regression model for the considered region of factor space and the isochoric model has become a winner from the prediction point of view.

## Appendix A

$$var\begin{bmatrix}\mathbf{b}_1\\ \mathbf{b}_2\end{bmatrix}=\begin{bmatrix}\left(\mathbf{X}^T\mathbf{X}\right)^{-1}+\left(\mathbf{X}^T\mathbf{X}\right)^{-1}\mathbf{X}^T\mathbf{Y}\mathbf{Q}^{-}\mathbf{Y}^T\mathbf{X}\left(\mathbf{X}^T\mathbf{X}\right)^{-1} & -\left(\mathbf{X}^T\mathbf{X}\right)^{-1}\mathbf{X}^T\mathbf{Y}\mathbf{Q}^{-}\\ -\mathbf{Q}^{-}\mathbf{Y}^T\mathbf{X}\left(\mathbf{X}^T\mathbf{X}\right)^{-1} & \mathbf{Q}^{-}\end{bmatrix}\begin{bmatrix}\mathbf{X}^T\mathbf{X} & \mathbf{X}^T\mathbf{Y}\\ \mathbf{Y}^T\mathbf{X} & \mathbf{Y}^T\mathbf{Y}\end{bmatrix}$$

$$\begin{bmatrix}\left(\mathbf{X}^T\mathbf{X}\right)^{-1}+\left(\mathbf{X}^T\mathbf{X}\right)^{-1}\mathbf{X}^T\mathbf{Y}\mathbf{Q}^{-}\mathbf{Y}^T\mathbf{X}\left(\mathbf{X}^T\mathbf{X}\right)^{-1} & -\left(\mathbf{X}^T\mathbf{X}\right)^{-1}\mathbf{X}^T\mathbf{Y}\mathbf{Q}^{-}\\ -\mathbf{Q}^{-}\mathbf{Y}^T\mathbf{X}\left(\mathbf{X}^T\mathbf{X}\right)^{-1} & \mathbf{Q}^{-}\end{bmatrix}^T\sigma^2$$

$$=\begin{bmatrix}\mathbf{I} & \left(\mathbf{X}^T\mathbf{X}\right)^{-1}\mathbf{X}^T\left[\mathbf{I}-\mathbf{Y}\mathbf{Q}^{-}\mathbf{Y}^T\left\{\mathbf{I}-\mathbf{X}\left(\mathbf{X}^T\mathbf{X}\right)^{-1}\mathbf{X}^T\right\}\right]\mathbf{Y}\\ \mathbf{0} & \mathbf{Q}^{-}\mathbf{Y}^T\left\{\mathbf{I}-\mathbf{X}\left(\mathbf{X}^T\mathbf{X}\right)^{-1}\mathbf{X}^T\right\}\mathbf{Y}\end{bmatrix}$$

$$\begin{bmatrix}\left(\mathbf{X}^T\mathbf{X}\right)^{-1}+\left(\mathbf{X}^T\mathbf{X}\right)^{-1}\mathbf{X}^T\mathbf{Y}\mathbf{Q}^{-}\mathbf{Y}^T\mathbf{X}\left(\mathbf{X}^T\mathbf{X}\right)^{-1} & -\mathbf{Q}^{-}\mathbf{Y}^T\mathbf{X}\left(\mathbf{X}^T\mathbf{X}\right)^{-1}\\ -\left(\mathbf{X}^T\mathbf{X}\right)^{-1}\mathbf{X}^T\mathbf{Y}\mathbf{Q}^{-} & \mathbf{Q}^{-}\end{bmatrix}\sigma^2$$

$$=\begin{bmatrix}\mathbf{G}^{-1}\left\{\mathbf{I}+\mathbf{X}^{\mathrm{T}}\mathbf{Y}\mathbf{Q}^{-}\mathbf{Y}\mathbf{X}\mathbf{G}^{-1}-\mathbf{X}^{\mathrm{T}}\left(\mathbf{I}-\mathbf{Y}\mathbf{Q}^{-}\mathbf{Z}^{\mathrm{T}}\right)\mathbf{Y}\mathbf{G}^{-1}\mathbf{X}^{\mathrm{T}}\mathbf{Y}\mathbf{Q}^{-}\right\} & -\mathbf{Q}^{-}\mathbf{Y}^{\mathrm{T}}\mathbf{X}\mathbf{G}^{-1}+\mathbf{G}^{-1}\mathbf{X}^{\mathrm{T}}\left(\mathbf{I}-\mathbf{Y}\mathbf{Q}^{-}\mathbf{Z}^{\mathrm{T}}\right)\mathbf{Y}\mathbf{Q}^{-}\\ -\mathbf{Q}^{-}\mathbf{Z}^{\mathrm{T}}\mathbf{Y}\mathbf{G}^{-1}\mathbf{X}^{\mathrm{T}}\mathbf{Y}\mathbf{Q}^{-} & \mathbf{Q}^{-}\mathbf{Z}^{\mathrm{T}}\mathbf{Y}\mathbf{Q}^{-}\end{bmatrix}\sigma^2$$

## Appendix B

$$var(\hat{\mathbf{y}})=[\mathbf{X}\ \mathbf{Y}]\begin{bmatrix}\mathbf{X}^T\mathbf{X} & \mathbf{X}^T\mathbf{Y}\\ \mathbf{Y}^T\mathbf{X} & \mathbf{Y}^T\mathbf{Y}\end{bmatrix}^{-}\begin{bmatrix}\mathbf{X}^T\\ \mathbf{Y}^T\end{bmatrix}\sigma^2$$

$$=[\mathbf{X}\ \mathbf{Y}]\begin{bmatrix}\left(\mathbf{X}^T\mathbf{X}\right)^{-1}+\left(\mathbf{X}^T\mathbf{X}\right)^{-1}\mathbf{X}^T\mathbf{Y}\mathbf{Q}^{-}\mathbf{Y}^T\mathbf{X}\left(\mathbf{X}^T\mathbf{X}\right)^{-1} & -\left(\mathbf{X}^T\mathbf{X}\right)^{-1}\mathbf{X}^T\mathbf{Y}\mathbf{Q}^{-}\\ -\mathbf{Q}^{-}\mathbf{Y}^T\mathbf{X}\left(\mathbf{X}^T\mathbf{X}\right)^{-1} & \mathbf{Q}^{-}\end{bmatrix}\begin{bmatrix}\mathbf{X}^T\\ \mathbf{Y}^T\end{bmatrix}\sigma^2$$

$$=[\mathbf{X}\ \mathbf{Y}]\begin{bmatrix}\left(\mathbf{X}^T\mathbf{X}\right)^{-1}\mathbf{X}^T+\left(\mathbf{X}^T\mathbf{X}\right)^{-1}\mathbf{X}^T\mathbf{Y}\mathbf{Q}^{-}\mathbf{Y}^T\mathbf{X}\left(\mathbf{X}^T\mathbf{X}\right)^{-1}\mathbf{X}^T & -\left(\mathbf{X}^T\mathbf{X}\right)^{-1}\mathbf{X}^T\mathbf{Y}\mathbf{Q}^{-}\mathbf{Y}^T\\ -\mathbf{Q}^{-}\mathbf{Y}^T\mathbf{X}\left(\mathbf{X}^T\mathbf{X}\right)^{-1}\mathbf{X}^T & \mathbf{Q}^{-}\mathbf{Y}^T\end{bmatrix}\sigma^2$$

$$=[\mathbf{X}(\mathbf{X}^{\mathrm{T}}\mathbf{X})^{-1}\mathbf{X}^{\mathrm{T}}\{\mathbf{I}+\mathbf{Y}\mathbf{Q}^{-}\mathbf{Y}^{\mathrm{T}}\mathbf{X}(\mathbf{X}^{\mathrm{T}}\mathbf{X})^{-1}\mathbf{X}^{\mathrm{T}}-\mathbf{Y}\mathbf{Q}^{-}\mathbf{Y}^{\mathrm{T}}\}+\mathbf{Y}\mathbf{Q}^{-}\mathbf{Y}^{\mathrm{T}}\{\mathbf{I}-\mathbf{X}(\mathbf{X}^{\mathrm{T}}\mathbf{X})^{-1}\mathbf{X}^{\mathrm{T}}\}]\,\sigma^2$$

$$=[\mathbf{X}(\mathbf{X}^{\mathrm{T}}\mathbf{X})^{-1}\mathbf{X}^{\mathrm{T}}+\mathbf{X}(\mathbf{X}^{\mathrm{T}}\mathbf{X})^{-1}\mathbf{X}^{\mathrm{T}}\mathbf{Y}\mathbf{Q}^{-}\mathbf{Y}^{\mathrm{T}}\mathbf{X}(\mathbf{X}^{\mathrm{T}}\mathbf{X})^{-1}\mathbf{X}^{\mathrm{T}}-\mathbf{X}(\mathbf{X}^{\mathrm{T}}\mathbf{X})^{-1}\mathbf{X}^{\mathrm{T}}\mathbf{Y}\mathbf{Q}^{-}\mathbf{Y}^{\mathrm{T}}+\mathbf{Y}\mathbf{Q}^{-}\mathbf{Y}^{\mathrm{T}}\{\mathbf{I}-\mathbf{X}(\mathbf{X}^{\mathrm{T}}\mathbf{X})^{-1}\mathbf{X}^{\mathrm{T}}\}]\,\sigma^2$$

$$=[\mathbf{X}(\mathbf{X}^{\mathrm{T}}\mathbf{X})^{-1}\mathbf{X}^{\mathrm{T}}-\mathbf{X}(\mathbf{X}^{\mathrm{T}}\mathbf{X})^{-1}\mathbf{X}^{\mathrm{T}}\mathbf{Y}\mathbf{Q}^{-}\mathbf{Y}^{\mathrm{T}}\{\mathbf{I}-\mathbf{X}(\mathbf{X}^{\mathrm{T}}\mathbf{X})^{-1}\mathbf{X}^{\mathrm{T}}\}+\mathbf{Y}\mathbf{Q}^{-}\mathbf{Z}^{\mathrm{T}}]\,\sigma^2$$

$$=\{\mathbf{X}(\mathbf{X}^{\mathrm{T}}\mathbf{X})^{-1}\mathbf{X}^{\mathrm{T}}-\mathbf{X}(\mathbf{X}^{\mathrm{T}}\mathbf{X})^{-1}\mathbf{X}^{\mathrm{T}}\mathbf{Y}\mathbf{Q}^{-}\mathbf{Z}^{\mathrm{T}}+\mathbf{Y}\mathbf{Q}^{-}\mathbf{Z}^{\mathrm{T}}\}\,\sigma^2$$

$$=[\mathbf{X}(\mathbf{X}^{\mathrm{T}}\mathbf{X})^{-1}\mathbf{X}^{\mathrm{T}}+\{\mathbf{I}-\mathbf{X}(\mathbf{X}^{\mathrm{T}}\mathbf{X})^{-1}\mathbf{X}^{\mathrm{T}}\}\mathbf{Y}\mathbf{Q}^{-}\mathbf{Z}^{\mathrm{T}}]\,\sigma^2$$

$$=\{\mathbf{X}(\mathbf{X}^{\mathrm{T}}\mathbf{X})^{-1}\mathbf{X}^{\mathrm{T}}+\mathbf{Z}(\mathbf{Z}^{\mathrm{T}}\mathbf{Z})^{-}\mathbf{Z}^{\mathrm{T}}\}\,\sigma^2$$